\documentclass[aps,twocolumn,pra,epsfig]{revtex4}

\usepackage{epsf}

\newcommand{\ba}{\begin{eqnarray}}
\newcommand{\ea}{\end{eqnarray}}
\newcommand{\bmath}{\begin{mathletters}}
\newcommand{\emath}{\end{mathletters}}
\newcommand{\ban}{\begin{eqnarray*}}
\newcommand{\ean}{\end{eqnarray*}}

\begin{document}

\title{ PROBABILITIES FROM ENTANGLEMENT, BORN'S RULE $p_k=|\psi_k|^2$ FROM ENVARIANCE}
\author{Wojciech Hubert Zurek}

    \address{Theory Division, MS B210, LANL
    Los Alamos, NM, 87545, U.S.A.}

\date{\today}

\begin{abstract}
I show how probabilities arise in quantum physics by exploring implications
of {\it environment - assisted invariance} or {\it envariance},
a recently discovered symmetry exhibited by entangled quantum systems.
Envariance of perfectly entangled ``Bell-like'' states can be used to rigorously justify
complete ignorance of the observer about the outcome of any measurement
on either of the members of the entangled pair. For more general states, envariance 
leads to Born's rule, $p_k \propto |\psi_k|^2$ for the outcomes associated with
Schmidt states. Probabilities derived in this manner are an
objective reflection of the underlying state of the system -- they represent
experimentally verifiable symmetries, and not just a subjective ``state of
knowledge'' of the observer. Envariance - based approach is compared with
and found superior to pre-quantum definitions of probability including
the {\it standard definition} based on the `principle of indifference' due
to Laplace, and the {\it relative frequency approach} advocated by von Mises.
Implications of envariance for the interpretation of quantum theory go beyond
the derivation of Born's rule: Envariance is enough to establish dynamical
independence of preferred branches of the evolving state vector of
the composite system, and, thus, to arrive at the {\it environment - induced
superselection (einselection) of pointer states}, that was usually derived
by an appeal to decoherence. Envariant origin of Born's rule for probabilities
sheds a new light on the relation between ignorance (and hence, information)
and the nature of quantum states.
\end{abstract}
\maketitle
\section{Introduction}

The aim of this paper is to derive Born's rule [1] and to identify and analyse
origins of probability and randomness in physics. The key idea we shall employ
is {\it environment - assisted invariance} (or {\it envariance}) [2-4],
a recently discovered quantum symmetry of entangled systems.  Envariance
allows one to use purity (perfect knowledge) of a joint state of an entangled pair 
to characterize unknown states of either of its components and to quantify
missing information about either member of the pair.

The setting of our discussion is essentially the same as in the study of 
decoherence and einselection [3,5-9]. However, the tools we shall
employ differ. Thus, as in decoherence, the system of interest ${\cal S}$ 
is ``open'' and can be entangled with its environment ${\cal E}$. We shall, 
however, refrain from using ``trace'' and ``reduced density matrix''.  Their
physical significance is based on Born's rule [10,11]. Therefore, to avoid 
circularity, we shall focus on pure global quantum states which yield -- 
as a consequence of envariance -- mixed states of their components. 
Successful derivation of Born's rule will in turn justify the usual interpretation 
of these formal tools while shedding a new light on the foundations of 
quantum theory and its relation with information.

The nature of  ``missing information'' and the origin of probabilities in
quantum physics are two related themes, closely tied to its interpretation.
We will be therefore forced to examine, in light of envariance, the structure
of the whole interpretational edifice. These fragments that depend on 
decoherence and einselection will have to be rebuilt without the standard 
tools such as trace and reduced density matrices. Once Born's rule is
``off limits'', the problem becomes not just to derive probabilities, and,
thus, to crown the already largely finished structure with $p_k=|\psi_k|^2$
as a final touch. Rather, the task is to deconstruct the interpretational
edifice standing on an incomplete, shaky foundation, and to re-build it using
these elements of the old plan that are still viable, but on a new, solid and
deep foundations and, to a large extent, from new, more basic building blocks.
We start in the next section with the proof that leads from envariance to
Born's rule. This will provide an overview of key ideas and their implications.

The original presentations of envariance [2-4] as well as most of this paper 
assume that quantum theory is universally valid. They rely only on unitary 
quantum evolutions and thus can be (as is decoherence) conveniently explored 
in the relative state framework [12] (although their results are independent of
interpretation). One may equally well -- as was emphasized by Howard Barnum [13] 
-- analyze envariance in a `Copenhagen setting' that includes the collapse postulate. 
Section II bypasses the discussion of quantum measurements and is in that sense
most explicitly interpretation - neutral study of the consequences of envariance.

The goal of this paper is to understand the nature of probabilities 
and to derive Born's rule in bare quantum theory. Thus, only unitary 
evolutions are allowed. The effective collapse (usually modeled 
with the help decoherence, which is in effect  ``off limits'' here) is all
one can hope for. The ground for the solution of the problem of the emergence
of probabilities in this setting is explored and prepared in Section III.
We start by comparing envariant definition of probability with the approaches
used in classical physics, and go on to examine quantum measurements.
``Symptoms of the classical''  (such as the preferred basis) that are taken 
for granted in justifying the need for probabilities and are usually derived with
the help of the trace operation and reduced density matrices are pointed out. 
As these tools depend on Born's rule, their physical implications such as 
decoherence and einselection have to be re-examined and often re-derived 
if we are to avoid circularity, and we set the stage for this in Section III.

Preferred pointer states are such a necessary pre-condition for the emergence
of the classical. Pointer states define what information is missing -- what are 
the potential measurement outcomes, providing `menu' of the alternative future 
events the observer may be ignorant of. Hence, they are indispensable in defining
probabilities. Pointer states are recovered in Section IV without decoherence
-- without relying on tools that implicitly invoke Born's rule: We show how
environment - induced superselection can be understood through direct appeal 
to the nature of the quantum correlations, and, in particular, to envariance. 
This pivotal result -- in a sense `einselection without decoherence' -- allows us 
to avoid any circularity in the discussion. It is based on the analysis
of correlations between the system and the apparatus pointer (or the memory
of the observer) in presence of the environment. It allows one to define future
events --  ``buds'' of the dynamically independent branches that can be assigned
probabilities.

Section V discusses probabilities from the ``personal point of view'' of 
an observer described by quantum theory.  Probabilities arise when 
the outcome of a measurement that is about to be performed cannot 
be predicted with certainty from the available data, even though observer 
knows all that can be known -- the initial pure state of the system.  

Relative frequencies of outcomes are considered
in the light of envariance in section VI, shedding a new light on
the connection between envariance and the statistical implications
of quantum states.

In course of the analysis we shall discover that none of the standard classical
approaches to probability apply directly in quantum theory. In a sense,  common
statement of the goal -- ``recovering classical probabilities in the quantum
setting'' -- may have been the key obstacle in making progress because it
was not ambitious enough. To be sure, it was understood long ago that none of
the traditional approaches to the definition of probability in the classical
world were all that convincing: They were either too subjective (relying
on the analysis of observers'  ``state of mind'', his lack of knowledge 
about the actual state) or too artificial (requiring infinite ensembles).

Complementarity of quantum theory provides the `missing ingredient' that
allows us to define probabilities using {\it objective properties} of entangled 
quantum states: Observer can know completely a global pure state of 
a composite system. That global state will have objective symmetries: They can be
experimentally verified and confirmed using transformations and measurements
that yield outcomes with certainty (and, hence, that do not involve Born's rule).
These objective properties of the global state imply -- as we shall see --
probabilities for the states of  local subsystems. Perfect information about the whole
can be thus used to demonstrate and quantify ignorance about a part. Circularity
of classical approaches (which assume ignorance -- e.g., ``equal likelihood'' -- to
establish ignorance -- probabilities) is avoided: Probabilities enter as an objective
property of a state. They reflect perfect knowledge (rather than ignorance) of 
the observer. These and other interpretational issues are discussed
in Section VII.

This paper can be read either in the order of presentation, or `like an onion',
starting from the outer layers (Sections II and VII, followed by III and IV, etc.)

\section{Probabilities from envariance}

To derive Born's rule we recognize that;
\begin{description}
\item[(o)] the Universe consists of systems;
\item[(i)] a completely known (pure) state of system ${\cal S}$ can
be represented by  a normalized vector in its Hilbert space ${\cal H_S}$;
\item[(ii)] a composite pure state of several systems is a vector
in the tensor product of the constituent Hilbert spaces;
\item[(iii)] states evolve in accord with the Schr\"odinger
equation $i \hbar |\dot \psi \rangle = H |\psi\rangle$ where $H$ is hermitean.
\end{description}

In other words, we start with the usual assumptions of the `no collapse' part
of quantum mechanics. We have listed them here in a somewhat
more `fine-grained' manner than it is often seen, e.g. in Ref. [3].

\subsection{Environment - assisted invariance}

Envariance is a symmetry of composite quantum states:
When a state $|\psi_{\cal SE}\rangle$ of a pair of systems ${\cal S,~E}$
can be transformed by  $U_{\cal S}=u_{\cal S} \otimes {\bf 1}_{\cal E}$ acting
solely on ${\cal S}$, 
$$ U_{\cal S}|\psi_{\cal SE}\rangle  =
(u_{\cal S} \otimes {\bf 1}_{\cal E})|\psi_{\cal SE}\rangle  =
|\eta_{\cal SE}\rangle \ ,  \eqno(1a)$$
but the effect of $U_{\cal S}$ can be undone by acting
solely on ${\cal E}$ with an appropriately chosen $U_{\cal E}=
{\bf 1}_{\cal S} \otimes u_{\cal E}$: 
$$U_{\cal E}|\eta_{\cal SE}\rangle  =
({\bf 1}_{\cal S} \otimes u_{\cal E}) |\eta_{\cal SE}\rangle
= |\psi_{\cal SE}\rangle \  , \eqno(1b)$$
then $|\psi_{\cal SE}\rangle$ is called envariant under $U_{\cal S}$.

In contrast to the usual symmetries (that describe situations when the action
of some transformation has no effect on some object) envariance is
an {\it assisted symmetry}: The global state is transformed by $U_{\cal S}$,
but can be restored by acting on ${\cal E}$, some other subsystem of
the Universe, physically distinct (e.g., spatially separated) from ${\cal S}$.
We shall call the part of the global state that can be acted upon to affect
such a restoration of the preexisting global state {\it the environment}
${\cal E}$.  Hence, the {\it environment - assisted invariance}, or -- for brevity 
-- {\it envariance}. We shall soon see that there may be more than one
such subsystem. In that case we shall use ${\cal E}$ to designate their 
union. Moreover, on occasion we shall consider manipulating
or measuring ${\cal E}$. So the oft repeated (and largely unjustified; see
Refs. [3,4,8,14,15]) phrase `inaccessible environment' should not be taken for
granted here.

Envariance of pure states is a purely quantum symmetry: Classical
state of a composite system is given by Cartesian (rather than tensor)
product of its constituents. So, to completely know the state of a composite
classical system one must know the state of each of its parts. It follows that
when one part of a classical composite system is affected by the analogue of
$U_{\cal S}$, the ``damage"  cannot be undone -- the state of the whole cannot
be restored -- by acting on some other part of the whole. Hence, pure classical
states are never envariant.

Another way of stating this conclusion is to note that states of classical
objects are `absolute', while in quantum theory there are situations
-- entanglement -- in which states are {\it relative}. That is, in classical
physics one would need to `adjust' the remainder of the universe to exhibit
envariance, while in quantum physics it suffices to act on systems entangled
with the system of interest. For instance, in a hypothetical classical universe
containing two and only two objects, a boost applied to either object could be
countered by simultaneously applying the same boost to the other: The
only motion in such a two - object universe is relative. Therefore, simultaneous
boosts would make the new state of that hypothetical universe
indistingushable from (and, hence, identical to) the initial pre-boost state.
This will not work in our Universe, as the center of mass of the two boosted
objects will be now moving with respect to the rest of its matter content. (That
is, unless we make the second object the rest of our Universe: this thought
experiment brings to mind the famous `Newton's bucket' -- i.e., Newton's
suspicion that the meniscus formed by water rotating in a bucket would
disappear if the rest of the Universe was forced to co-rotate.)

To give an example of envariance, consider Schmidt decomposition of
$\psi_{\cal SE}$;
$$|\psi_{\cal SE}\rangle = \sum_{k=1}^N a_k |s_k\rangle |\varepsilon_k\rangle
\ . \eqno(2)$$
Above, by definition of Schmidt decomposition, $\{|s_k\rangle\}$
and $\{|\varepsilon_k\rangle\}$ are orthonormal and $a_k$ are complex.
Any pure bipartite state can be written this way.
A whole class of envariant transformations can be identified
for such pure entangled quantum states:

\noindent{\bf Lemma 1.} Any unitary transformation with Schmidt eigenstates
$\{|s_k\rangle\}$;
$$u_{\cal S}=\sum_{k=1}^N \exp(i \phi_k)|s_k\rangle \langle s_k| \ , 
\eqno(3a)$$
is envariant.

\noindent{\bf Proof}: Indeed, any unitary with Schmidt eigenstates 
can be undone by
a `countertransformation':
$$u_{\cal E}=\sum_{k=1}^N \exp(-i \phi_k + 2 \pi l_k)|\varepsilon_k\rangle
\langle \varepsilon_k|  \eqno(3b)$$
where $l_k$ are arbitrary integers. QED.

\noindent {\bf Remark}: The environment used to undo
$u_{\cal S}$ need not be uniquely defined: For example, $u_{\cal S}$ acting
on a GHZ-like state:
$$|\psi_{{\cal SEE}'}\rangle = \sum_{k=1}^N a_k |s_k\rangle
|\varepsilon_k\rangle |\varepsilon_k'\rangle \eqno(4a) $$
can be envariantly undone by acting either on ${\cal E}$ or on ${\cal E}'$,
or by acting on both parts of the joint environment.

It is perhaps useful to point out that one can use $\psi_{\cal SEE'}$ to
obtain reduced density matrix:
$$\rho_{\cal SE} = \sum_k |a_k|^2 |s_k\rangle \langle s_k | \otimes
| \varepsilon_k \rangle \langle \varepsilon_k| \eqno(4b)$$
This means that even when the correlated state of ${\cal S}$ and ${\cal E}$
is mixed and of this form, one can in principle imagine that there is 
an underlying pure state. States of the above form can arise in measurements
or as a consequence of decoherence. The discussion can be thus re-phrased 
in terms of pure states in all cases of interest. The assumption of suitable
pure global state entails no loss of generality.

At first sight, envariance may not seem to be all that significant,
since it is possible to show that it can affect only phases:

\noindent{\bf Lemma 2}. All envariant unitary transformations have eigenstates
that coincide with the Schmidt expansion of $|\psi_{\cal SE}\rangle$, i.e.,
have the form of Eq. (3a).

\noindent{\bf Proof} is by contradiction: Suppose there is an envariant unitary
$\tilde u_{\cal S}$ that cannot be made co-diagonal with Schmidt basis of
$\psi_{\cal SE}$, Eq. (2). It will then inevitably transform Schmidt states
of ${\cal S}$:
$$\tilde u_{\cal S} \otimes {\bf 1}_{\cal E} |\psi_{\cal SE}\rangle =
\sum_{k=1}^N a_k (\tilde u_{\cal S} |s_k\rangle) |\varepsilon_k\rangle =
\sum_{k=1}^N a_k |\tilde s_k\rangle |\varepsilon_k\rangle =
|\tilde \eta_{\cal SE}\rangle  \ .$$
If $\tilde u_{\cal S}$ is envariant there must be $\tilde u_{\cal E}$ such that
$${\bf 1}_{\cal E} \otimes \tilde u_{\cal E} |\tilde \eta_{\cal SE}\rangle =
\sum_{k=1}^N a_k |s_k\rangle |\varepsilon_k\rangle =
|\psi_{\cal SE}\rangle \ . $$
But unitary transformations acting exclusively on ${\cal H_E}$ cannot change
states in ${\cal H_S}$. So, the new set of Schmidt states $|\tilde s_k\rangle$
of $\tilde \eta_{\cal SE}$ cannot be undone (`rotated back') to $|s_k\rangle$  
by any $\tilde u_{\cal E}$. It follows that -- when Schmidt  states are uniquely
defined -- there can be no envariant unitary transformation that acts on
the environment and restores the global state to $\psi_{\cal SE}$ after
Schmidt states of ${\cal S}$ were altered by $\tilde u_{\cal S}$. QED.

\noindent{\bf Corollary:} Properties of global states are envariant iff
they are a function of the phases of the Schmidt coefficients.

Phases are often regarded as inaccessible, and are even sometimes dismissed 
as unimportant (textbooks tend to speak of a `ray' in the Hilbert space, thus 
defining a state modulo its phase). Indeed, Schmidt expansion is occasionally 
defined by absorbing phases in the states which means that all the non-zero 
coefficients end up real and positive (and hence all the phases are taken to be zero).
This is a dangerous oversimplification. Phases matter -- reader can verify that 
it is impossible to write all of the Bell states when all the relative phases are set 
to zero. Indeed, the aim of the rest of this paper is, in a sense, to carefully justify
when and for what purpose phases can be disregarded, and to understand 
nature of the ignorance about the local state of the system as a consequence
of the global nature of these phases.  

\subsection{State of a subsystem of a quantum system}

Independence of the state of the system ${\cal S}$ from phases of
the Schmidt coefficients will be our first important conclusion based 
on envariance. To establish it we list below three {\bf facts} -- additional
assumptions that may be regarded as obvious.  We state them here
explicitly to clarify and extend the meaning of terms `(sub)system' and `state' 
we have already used in axioms (o) - (iii).
\begin{description}
\item[{\bf Fact 1}:] Unitary transformations must act on the system to alter
its state. (That is, when the evolution operator does not operate on the
Hilbert space ${\cal H_S}$ of the system, i.e., when it has a form $...\otimes
{\bf 1}_{\cal S} \otimes...$ the state of ${\cal S}$ remains the same.)
\item[{\bf Fact 2}:] The state of the system ${\cal S}$ is all that is needed
(and all that is available) to predict measurement outcomes, including their
probabilities.
\item[{\bf Fact 3}:] The state of a larger composite system that includes
${\cal S}$ as a subsystem is all that is needed (and all that is available)
to determine the state of the system ${\cal S}$.
\end{description}
\noindent We have already implicitly appealed to fact 1 earlier,
e.g. in the proof of Lemma 2. Note that the above {\bf facts} are interpretation
- neutral and that states (e.g., `the state of ${\cal S}$') they refer to need
not be pure.

With the help of the facts can now establish:

\noindent{\bf Theorem 1.} For an entangled global state of the system and the
environment all measurable properties of ${\cal S}$ -- including probabilities
of various outcomes -- cannot depend on the phases of Schmidt coefficients:
The state of ${\cal S}$ has to be completely determined by the set of pairs
$\{|\alpha_k|,~|s_k\rangle\}$.

\noindent{\bf Proof:} Envariant transformation $u_{\cal S}$ could affect
the state of ${\cal S}$. However, by definition of envariance the effect of
$u_{\cal S}$ can be undone by a countertransformation of the form
${\bf 1}_{\cal S} \otimes u_{\cal E}$ which -- by fact 1 -- cannot alter
the state of ${\cal S}$. As ${\cal SE}$ is returned to the initial
state, it follows from fact 3 that the state of ${\cal S}$ must have been
also restored. But (by fact 1) it could not have been effected by the
countertransformation. So it must have been left unchanged by the envariant
$u_{\cal S}$ in the first place. It follows (from the above and (fact 2)
that measurable properties of ${\cal S}$ are unaffected by envariant
transformations. But, by Lemma 1 \& 2, envariant transformations can alter
phases and only phases of Schmidt coefficients. Therefore, any measurable
property of ${\cal S}$ implied by its state must indeed be completely 
determined by the set of pairs $\{|\alpha_k|,~|s_k\rangle\}$. QED.

\noindent {\bf Remark}: Information content of the list
$\{|\alpha_k|,~|s_k\rangle\}$ that describes the state of ${\cal S}$
is the same as the information content of the reduced density matrix.
We do not know yet, however, what are the probabilities of various outcome
states $|s_k\rangle$.

Thus, envariance of Schmidt phases proves that only absolute values 
of Schmidt coefficients can influence 
measurement
outcomes. Yet, the dismissive attitude towards phases we have reported above
is incorrect. This is best illustrated on an example: Changing phases between
the Hadamard states;
$$ |\pm\rangle = (|s_1\rangle \pm |s_2\rangle)/\sqrt 2 $$
can change the state of the system from $|s_1\rangle$ to $|s_2\rangle$.
More generally;

\noindent{\bf Lemma 3}. Iff the Schmidt decomposition of Eq. (2) has 
coefficients
that have the same absolute value -- that is, the state is {\it even}:
$$|\bar \psi_{\cal SE}\rangle \propto \sum_{k=1}^N e^{i \phi_k} 
|s_k\rangle |\varepsilon_k\rangle  \ , \eqno(5)$$
it is also envariant under a {\it swap}:
$$ u_{\cal S}(1 \rightleftharpoons 2 ) =
e^{i \phi_{12}} |s_1\rangle \langle s_2|  + h.c. \eqno(6a)$$

\noindent{\bf Proof}: By Lemma 1, swap is envariant -- it can be generated by
$u_{\cal S}$ diagonal in Hadamard basis of the two states (which is 
also Schmidt when their coefficients differ only by a phase). Swaps can be seen 
to be envariant also more directly: When $a_1=|a|e^{i \phi_1},\ 
a_2=|a|e^{i \phi_2}$,
every swap can be undone by the corresponding {\it counterswap}:
$$ u_{\cal E}(1 \rightleftharpoons 2) = e^{-i( \phi_{12} + \phi_1 - \phi_2  +
2 \pi l_{12} ) }|\varepsilon_1\rangle \langle \varepsilon_2| + h.c. \eqno(6b)$$
This proves envariance of swaps for equal values of the coefficients of the
swapped states. Converse follows from Lemma 1 \& 2: Envariant transformations
can affect only {\it phases} of Schmidt coefficients, so the global 
state cannot
be restored after the swap when their {\it absolute values} differ. QED.

\noindent{\bf Remark}: When ${\cal SE}$ is in an even state
$|\bar \psi_{\cal SE}\rangle$ (that is therefore envariant under swaps)
exchange $|s_1\rangle \rightleftharpoons |s_2\rangle$ does not affect
the state of ${\cal S}$ -- its consequences cannot be detected by any
measurement of ${\cal S}$ alone.

Lemma 3 we have just established is the cornerstone of our approach. We now
know that when the global state of ${\cal SE}$ is even (i.e., with equal
absolute values of its Schmidt coefficients), then a swap (which predictably
takes {\it known} $|s_k\rangle$ into $|s_l\rangle$) does not alter the state
of ${\cal S}$ at all. 

{\it Envariantly swappable state of the system defines perfect ignorance}. 
We emphasize the direction of this implication: The state of ${\cal S}$ is 
provably completely unknown {\it not} because of the subjective ignorance 
of the oberver. Rather, it is unknown as a consequence of complementarity: 
the state of ${\cal SE}$ is after all perfectly known.  Moreover, it can 
be objectively known to many observers. All of them will agree that their 
perfect global knowledge implies complete local ignorance. Therefore, 
{\it probabilities are objective properties of this state}. 

Symmetries of the state of ${\cal SE}$ imply ignorance of the observer 
about the outcomes of his future measurements on ${\cal S}$. This emergence
of objective probabilities is purely quantum. Objective probabilities are 
incompatible with classical setting (where there is an unknown but definite 
preexisting state). In the quantum setting, objective nature of probabilities 
arises as a consequence of the entanglement, e.g., with the environment.

\subsection{Born's rule from envariance}

So far, we have avoided refering to probabilities. Apart from brief mention
in fact 2 and immediately above we have not discussed how do they relate to
quantum measurements. This will have to wait for when we consider quantum
measurements, records and observers from an envariant point of view. But it
turns out that one can derive the rule connecting probabilities with entangled
state vectors such as $|\psi_{\cal SE}\rangle$ of Eq. (2) from 
relatively modest
assumptions about their properties. The next key step in this direction is:

\noindent{\bf Theorem 2.} Probabilities of Schmidt states of ${\cal S}$ that
appear in $|\psi_{\cal SE}\rangle$ with coefficients that have same absolute
value are equal.

There are several inequivalent ways to establish Theorem 2. Indeed, the reader
may feel that it was already established: Remark that followed Lemma 3 plus
a rudimentary symmetry arguments suffice to do just that. However,
for completeness, we now spell out some of these arguments in more detail.
Both Barnum [13] and Schlosshauer and Fine [16] have discussed some of the
related issues. Reporting some of their conclusions (and anticipating some
of ours) it appears that envariance plus a variety of small subsets of natural
assumptions suffice to arrive at the thesis of Theorem 2.

\subsubsection{Envariance under complete swaps}

We start with the first version of the proof:

\begin{description}
\item[(a)] {\it When operations that swap any two orthonormal states leave
the state of ${\cal S}$ unchanged, probabilities of the outcomes associated
with these states are equal.}
\end{description}
\noindent{\bf Proof (a)} is immediate. When the entangled state of ${\cal SE}$
has equal values of the Schmidt coefficients, e.g. $|\bar \psi_{\cal SE}\rangle
\propto \sum_{k=1}^N e^{i \phi_k} |s_k\rangle |\varepsilon_k\rangle $,
local state of ${\cal S}$ will be indeed unaffected by the swaps (by
Theorem 1 and Lemma 3 above). Consequently, with the assumption (a), thesis
of Theorem 2 follows. QED.

The above argument is close in the spirit to Laplace's ``principle of
indifference" [17]. We prove that swapping possible outcome states -- shuffling
cards -- ``makes no difference''. However, in contrast to Laplace {\it we show
`objective indifference' of the physical state of the system in question rather 
than observers subjective indifference based on his state of knowledge}.

Note that in absence of entanglement (and, hence envariance) swap generally 
changes the underlying state of the system also when coefficients of states
corresponding to various potential outcomes have the same absolute values. 
For example, pure states 
$|\alpha\rangle \propto |1\rangle +|2\rangle-|3\rangle+|4\rangle$ and
$|\beta\rangle \propto |1\rangle +|3\rangle-|2\rangle+|4\rangle$ are orthogonal
even though they have the same absolute values of the coefficients and differ
only by a swap. Thus, without entanglement with the environment (i.e.,
in absence of Lemma 1 \& 2 and, hence, Theorem 1 that allows one to
ignore phases of Schmidt coefficients) assumption (a) would be tantamount
to assertion that phases of the coefficients are unimportant in specifying
the state. For isolated systems this is obviously wrong, in blatant 
conflict with the quantum principle of superposition! 

In absence of envariance the key to our argument -- assertion that the 
{\it state} of ${\cal S}$ is left unchanged by a swap -- is simply wrong. 
This is easily seen by considering ensemble of identical pure states (such
as $|\alpha\rangle$). Through measurements, observer can
find out the state of systems in that ensemble (e.g., 
 $|\alpha\rangle$ or $|\beta\rangle$). By contrast,
if this ensemble becomes first entangled with the environment in
such a way that $|1\rangle \dots |4\rangle$ are Schmidt, an observer with
access to ${\cal S}$ only would conclude that the state of the system is
a perfect mixture, and would not be able to tell if the pre-decoherence state
was $|\alpha\rangle$ or $|\beta\rangle$.

\subsubsection{Envariance under partial swaps and dynamics}

The second strategy we shall use to prove Theorem 2 relies on a somewhat
different `dynamical' definition of indifference. We note that when we have
some information about the state, we should be able to ``detect motion''
(possibly using an ensemble of such states) -- to observe changes caused by
the dynamical evolution. This intuition is captured by our assumption:
\begin{description}
\item[(b)] {\it When the state of ${\cal S}$ is left unchanged by all
conceivable unitary transformations acting on a subspace
$\tilde {\cal H}_{\cal S}$ of ${\cal H_S}$, then probabilities of all
the outcomes of any exhaustive measurement corresponding to any orthocomplete
basis that spans that subspace of $\tilde {\cal H}_{\cal S}$ are the same.}
\end{description}
To proceed we first establish that when complete swaps, Eq. (6a), between a
subset of states of some orthonormal basis that spans $\tilde {\cal 
H}_{\cal S}$
leave the state of ${\cal S}$ unchanged, then so do {\it partial swaps} on the same
subspace:

\noindent{\bf Lemma 4}: {\it Partial swaps} defined by pairwise exchange of
any two orthonormal basis sets $\{|s_k\rangle\}$ and $\{|\tilde s_l\rangle\}$
that span {\it even subspace} ${\bar {\cal H}_{\cal S}}$ of ${\cal H_S}$ which
admits full swaps, Eq. (6a) -- are also envariant.

\noindent{\bf Proof}: Partial swap can be expressed as a unitary;
$$ \tilde u_{\cal S}(\{ \tilde s_k\} \rightleftharpoons \{s_k\}) =
\sum_{|s_k\rangle \in \bar {\cal H_S}} |\tilde s_k \rangle \langle s_k|
\ . \eqno(6c)$$
This is the obvious generalization of the simple swap of Eq. (6a).
$ \tilde u_{\cal S}(\{ \tilde s_k\} \rightleftharpoons \{s_k\})$ can be undone
by the corresponding {\it partial counterswap} of the Schmidt partners of
the swapped pairs of states. It follows from Lemma 3 that $\bar \psi_{\cal SE}$
-- a state envariant under complete swaps -- must have a form:
$$|\bar \psi_{\cal SE}\rangle \propto \sum_{k=1}^K e^{i \phi_k}
|s_k\rangle |\varepsilon_k\rangle$$
where $K=Dim( \bar {\cal H_S})$. Basis $\{|\tilde s_l\rangle\}$ spans
the same subspace $\bar {\cal H}_{\cal S}$.  Therefore,
$$|\bar \psi_{\cal SE}\rangle \propto \sum_{l=1}^K |\tilde s_l \rangle
(\sum_{k=1}^K e^{i \phi_k} \langle \tilde s_l |s_k\rangle 
|\varepsilon_k\rangle)
= \sum_{l=1}^K |\tilde s_l\rangle |\tilde \varepsilon_l\rangle\ . $$
Given that $\{|\varepsilon_k\rangle\}$ are Schmidt, it is straightforward to
verify that $\{|\tilde \varepsilon_l\rangle\}$ are orthonormal, and, therefore,
the expansion on RHS above is also Schmidt. Consequently;
$$\tilde u_{\cal E}(\{\tilde\varepsilon_k\}\rightleftharpoons\{\varepsilon_k\})
= \sum_{\varepsilon_k \in \bar {\cal H_E}} |\tilde\varepsilon_k \rangle
\langle \varepsilon_k| \ , \eqno(6d)$$
the desired partial counterswap exists. This establishes envariance under
partial swaps. QED.

\noindent{\bf Corollary}: When complete swaps are envariant in the subspace
$\bar {\cal H_S} \in {\cal H_S}$, so are all the unitary transformations on
$\bar {\cal H_S}$. Indeed, the set of all partial swaps is the same
as the set of all unitary transformations on the subspace $\bar {\cal H_S}$.

We can now give the second proof of Theorem 2:

\noindent {\bf Proof (b)}: Equality of probabilities under envariant swaps
follows immediately from the above Corollary and assumption (b). QED.

Using Lemma 4 and its Corollary, we can identify mathematical objects that
represent even states in ${\cal H_S}$: Only a uniform  distribution of pure
states over $\bar {\cal H}_{\cal S}$ is invariant under all unitaries.
The alternative representation that is more familiar is the (reduced)
density matrix. It has to be proportional to the identity operator:
$$ \rho_{\cal S} \propto {\bf 1}_{\cal S} \ , $$
to be invariant under all unitaries.

So envariance, the no collapse axioms (o)-(iii), plus the three {\bf facts}
imply that our abstract state (whose role is defined by fact 2) leads to
the distribution uniform in the Haar measure or, equivalently to the reduced
density matrix $\propto{\bf 1}_{\cal S}$ within $\bar {\cal H}_{\cal S}$. Note
that these conclusions follow from Lemma 4, which does not employ assumption
(b). The form of the mathematical object representing envariantly swappable
state of ${\cal S}$ follows directly from the symmteries of the underlying
entangled composite state of ${\cal SE}$. In particular, we have in a sense
obtained the reduced density matrix in the special case without the usual
arguments [10,11], i.e., without relying on Born's rule. Moreover, assumption
(b) is needed only when we want to interpret that reduced density matrix
in terms of probabilities.

Assumption (b) can also be regarded as a quantum counterpart of
Laplace's {\it principle of indifference} [17]. Now there are however even
more obvious differences between the quantum situation and shuffling cards
than these we have already mentioned in the discussion of proof (a). Classical 
deck cannot be shuffled into a superposition of the original cards. This can 
obviously happen to a `quantum deck': In quantum physics we can 
consider arbitrary unitary transformations (and not just discrete
swaps). This consequence of the nature of quantum evolutions can be traced all
the way to the principle of superposition (and, hence, to phases!).  

The other distinction between the quantum and classical principle of indifference 
we have already noted is even more striking: In classical physics it was the
``state of knowledge" of the observer -- his description of the system -- that may 
(or may not) have been altered by the evolution -- the underlying physical state 
was {\it always} affected when shuffling / evolution was non-trivial.
In quantum theory there is no distinction between the epistemic `state of
knowledge' role of the state and its objective (`ontic') role. In this sense quantum 
states are `epiontic' [3]. 

Probabilities are -- in any case -- an {\it objective} reflection of symmetries of 
such states: They follow from quantum complementarity between the global
and local observables. They can be defined and quantified using envariance, 
an experimentally verifiable property of entangled quantum states.

\subsubsection{Equal probabilities from perfect ${\cal SE}$ correlations}

Both of the proofs above start with the assumption that under certain conditions
probabilities of a subset of states of the system are equal, and then establish
the thesis by showing that this assumption is implied by envariance under swaps
-- both are in that sense Laplacean. The third proof also starts with
an assumption of equality of probabilities, but now we consider relation
between the probabilities of the Schmidt states of ${\cal S}$ and ${\cal E}$.
This approach (Barnum [13], see also discussion in Ref. [16])
recognizes that pairs of Schmidt states ($|s_k\rangle|\varepsilon_k\rangle$
in Eq. (2)) are perfectly correlated, which implies that they have the same
probabilities. Thus, one can prove equality of probabilities
of envariantly swappable Schmidt states directly from envariance by relying
on perfect correlations between Schmidt states of ${\cal S}$ and ${\cal E}$.

To demonstrate this equality we consider a subspace of ${\cal H_S}$ spanned by
Schmidt states $|s_k\rangle$ and $|s_l\rangle$, so that the corresponding
fragment of $\psi_{\cal SE}$ is given by:
$$ |\psi_{\cal SE}\rangle = \dots + a_k|s_k\rangle|\varepsilon_k\rangle
+a_l|s_l\rangle|\varepsilon_l\rangle + \dots $$
We assume that:
\begin{description}
\item [(c)] {\it In the Schmidt decomposition partners are perfectly
correlated, i.e., detection of $|s_k\rangle$ implies that in a subsequent
measurement of Schmidt observable (i.e., observable with Schmidt eigenstates)
on ${\cal E}$ will certainly obtain $|\varepsilon_k\rangle$
(this `partner state' will be recorded with certainty, e.g., with probability 1).}
\end{description}
\noindent {\bf Proof (c)}: From assumption (c) we immediately have that
$$p_{\cal S}(s_{k(l)})= p_{\cal E} (\varepsilon_{k(l)}) \ . $$
Consider now a swap $s_k \rightleftharpoons s_l$. It leads from
$\psi_{\cal SE}$, Eq. (2), to:
$$|\eta_{\widetilde{\cal S}{\cal E}}\rangle=\dots+a_k|s_l\rangle
|\varepsilon_k\rangle +a_l|s_k\rangle|\varepsilon_l\rangle + \dots $$
Now, using again (c) we get;
$$p_{\widetilde {\cal S}} (s_{k(l)})
= p_{\cal E} (\varepsilon_{l(k)}) = p_{\cal S}(s_{l(k)})\ . $$
In effect, this establishes the `pedantic assumption' of [2,3] using envariance
and perfect correlation assumption (c): Probabilities get exchanged when
the states are swapped. So we could go back to the proof of Ref. [2] with
a somewhat different motivation. But we can also continue, and consider
a counterswap in ${\cal E}$ that yields:
$$|\vartheta_{\widetilde{\cal S}\widetilde{\cal 
E}}\rangle=\dots+a_k|s_l\rangle|\varepsilon_l\rangle
+a_l|s_k\rangle|\varepsilon_k\rangle + \dots $$
We now consider the case when $|a_k|=|a_l|=a$. Counterswap restores such
an even state;
$$|\bar \vartheta_{\widetilde{\cal S}\widetilde{\cal E}}\rangle
=\dots+a|s_l\rangle|\varepsilon_l\rangle
+a|s_k\rangle|\varepsilon_k\rangle + \dots = |\bar\psi_{\cal SE}\rangle \ . $$
By facts 2 and 3 the overall state as well as the state of ${\cal S}$ must have
been restored to the original. Therefore $ p_{\widetilde{\cal S}}(s_{l(k)}) =
p_{\cal S}(s_{l(k)})$. Together with $p_{\widetilde{\cal S}}(s_{l(k)})=
p_{\cal S}(s_{k(l)})$ established before this yields:
$$ p_{\cal S}(s_{l(k)}) = p_{\cal S}(s_{k(l)})  $$
when the relevant Schmidt coefficients have equal absolute values.
QED.

We have now established that in an entangled state in which all of
the coefficients have the same absolute value so that every state $|s_k\rangle$
can be envariantly swapped with every other state $|s_l\rangle$,
$|\bar\psi_{\cal SE}\rangle \propto \sum_{k=1}^N e^{i \phi_k} |s_k\rangle
|\varepsilon_k\rangle $, all the possible orthonormal outcome states
have the same probability. Let us also assume that states that
do not appear in the above superposition (i.e., appear with Schmidt
coefficient zero) have zero probability. (We shall motivate this rather natural
assumption later in the paper.) Given the customary normalisation
of probabilities we get for even (`Bell-like') states:

\noindent{\bf Corollary} For states with equal absolute values of Schmidt
coefficients;
$$ p_k = p (s_k) = 1/N \ \ \ \forall k \  . \eqno(7a)$$

Moreover, probability of any subset of $n$ mutually exclusive events
is additive. Hence:
$$ p_{k_1 \vee k_2 \vee \dots \vee k_n} =
p(s_{k_1} \vee s_{k_2} \vee \dots \vee s_{k_n} ) = n/N \ . \eqno(7b) $$
Above, we have assumed that orthogonal states correspond to mutually 
exclusive events. We shall motivate also this (very natural) assumption of 
the additivity of probabilities further in discussion of quantum measurements 
in Section V (thus going beyond the starting point of e.g. Gleason[30]). 
Here we only note that while additivity of probabilities looks innocent,
in the quantum case (where the principle of superposition entitles one to add
complex amplitudes) it should not be taken for granted. In the end, we shall
conclude that additivity of probabilities is tied to envariance, which yields
phases (and, hence, quantum superposition principle) irrelevant for 
Schmidt states of the subsystems of the entagled whole.

We also note that the probability entered our discussion in a manner that
bypasses circularity: We have simply identified certainty with the probability
of 1 (see e.g. assumption (c) above). This provides us with the normalization,
while the symmetries revealed by envariance determine probabilities
in the case when there is no certainty.

\subsection{Born's rule: the case of unequal coefficients}

To complete derivation of Born's rule consider the case when the absolute
values of the coefficients $a_k$ in the Schmidt decomposition are proportional
to $\sqrt{m_k}$, where $m_k$ are natural numbers.
$$|\psi_{\cal SE}\rangle \propto \sum_{k=1}^N \sqrt{m_k} e^{i \phi_k}
|s_k\rangle |\varepsilon_k\rangle \ . \eqno(8a)$$
We now introduce a {\it counterweight / counter} ${\cal C}$.  It can be thought
of either as a subsystem extracted from the environment ${\cal E}$, or as an
ancilla that becomes correlated with ${\cal E}$ so that the combined
state is:
$$|\psi_{\cal SEC}\rangle \propto \sum_{k=1}^N \sqrt{m_k} e^{i \phi_k}
|s_k\rangle |\varepsilon_k\rangle |C_k\rangle \ \eqno(8b)$$
where $\{|C_k\rangle\}$ are orthonormal. Moreover, we assume that $|C_k\rangle$
are associated with subspaces of ${\cal H_C}$ of sufficient dimensionality so
that the `fine-graining' represented by:
$$|C_k\rangle =  \sum_{j_k=\mu_{k-1}+1}^{\mu_k}
|c_{j_k}\rangle / \sqrt{m_{k}} \eqno(9a) $$
is possible. Above, $\mu_k=\mu_{k-1}+m_k$, and $\mu_0=0$.

We also utilize
a {\tt c-shift}, a {\tt c-not} - like gate[3] that correlates states
of ${\cal E}$ with the fine-grained states of ${\cal C}$:
$$|c_{j_k}\rangle |\varepsilon_k\rangle \longrightarrow |c_{j_k}\rangle
|e_{j_k}\rangle \ , \eqno(9b)$$
where $|e_{j_k}\rangle$ are orthonormal states of $\cal E$ that correlate with
the individual states of the counter $\cal C$ (e.g., causing decoherence).
We have now arrived at the state vector that represents a perfectly entangled
(equal coefficient, or `even') state of the composite system consisting of
${\cal SC}$ and ${\cal E}$:
$$|\Psi_{\cal SEC}\rangle \propto \sum_{j_k=1}^M e^{i \phi_{j_k}}
|s_{k(j_k)}, c_{j_k}\rangle |e_{j_k}\rangle \ . \eqno(8c)$$
Above, $M=\sum_{k=1}^N m_k = \mu_N$, and $k = k(j)$ is the obvious
``staircase'' function, i.e. when $\mu_{k-1} < j \le \mu_k$, then $k(j)=k$.

The state $\Psi_{\cal SEC}$ is envariant under swaps of joint Schmidt states
$|s_{k(j_k)}, c_{j_k}\rangle$ of ${\cal SC}$. Hence, by Eq. (7a):
$$ p_{j_k} \equiv p(c_{j_k}) \equiv  p(|s_{k(j_k)}, c_{j_k}\rangle ) = 1 / M$$
Moreover, Eq. (7b) implies that if we were to enquire about the
probability of the state of ${\cal S}$ alone, the answer must be given by:
$$p_k \ \equiv  \sum_{j_k=\mu_{k-1} + 1}^{\mu_{k}}
p(|s_{k(j_k)}, c_{j_k}\rangle )  =  m_k / M \  = \ |a_k|^2 \ . \eqno(10)$$
This is Born's rule. The extension to the case where $|a_k|^2$ are
incommensurate is straightforward by continuity as rational numbers
are dense among reals.

\section{Envariance, Ignorance, and Chance}

We have now presented a fairly complete discussion of envariance, 
and we have derived Born's rule. The aim of the rest of this paper is to consider
some of the other implications of envariance for our quantum Universe.
This section serves the role of the intermission after the first act of a play.
The basic plot is already in place. We can now take a few moments to
speculate on how it will develop. In particular, we shall compare definition 
of probabilities based on envariance with the pre-quantum discussions of this
concept.  We shall also set the stage for the investigation of the implications 
of envariance for the interpretation of quantum theory. This includes 
(but is not limited to) the issue of  the `decoherence-free' derivation 
of the preferred pointer basis.

\begin{figure*}[p]
\centering 
\epsfxsize 7in
\epsfbox{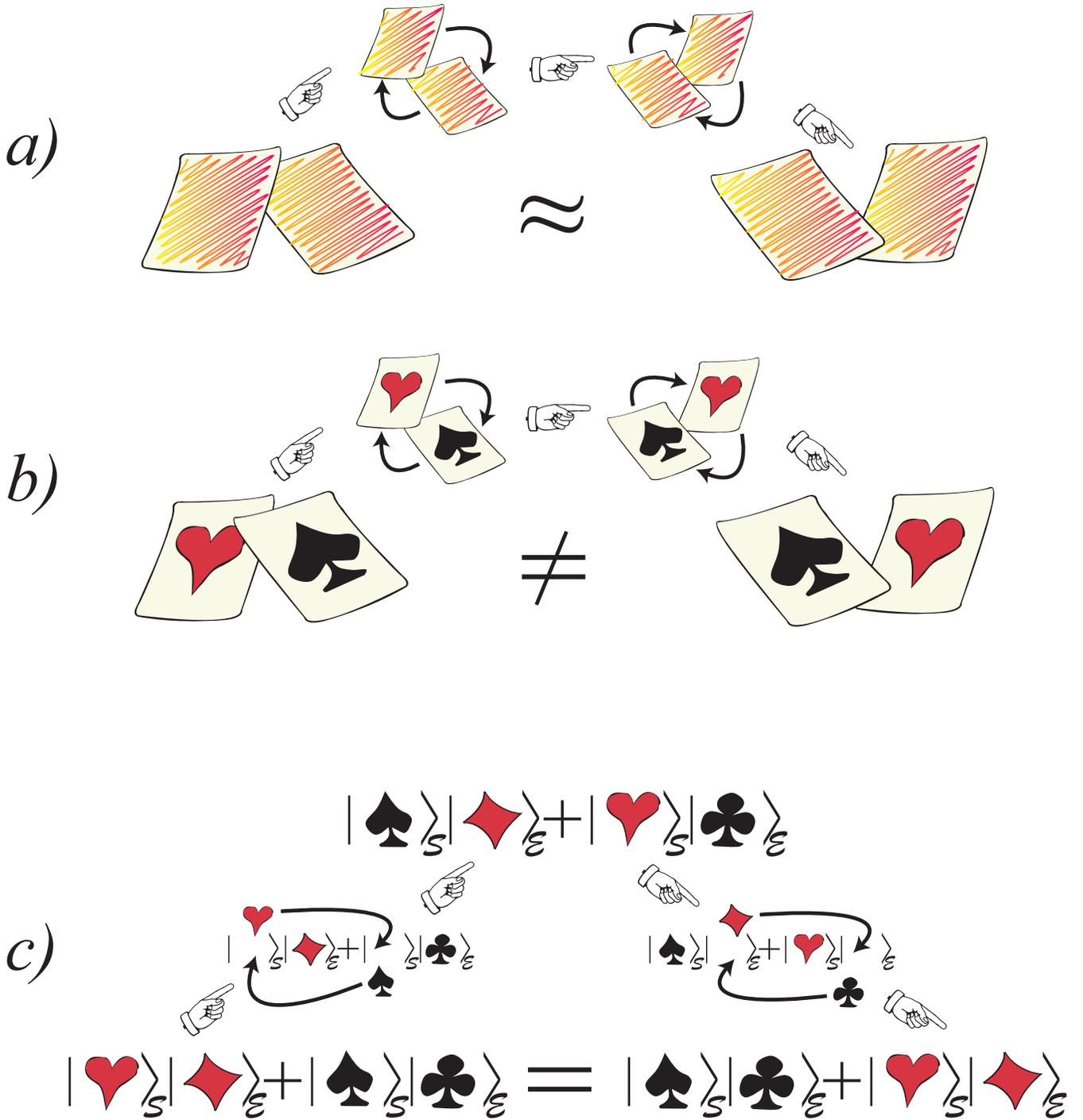}
\caption{Envariance is related to  the `principle of indifference' (or the `principle of
equal likelihood') used by Laplace [17] to define subjective probabilities. However
in quantum physics envariance leads to {\it probabilities based on an objective 
symmetries of the underlying physical state of the system}. The principle of indifference 
is illustrated in Fig. 1a. An observer (or a card player) who knows one of the two cards is
$\spadesuit$ but does not see their faces does not care -- is indifferent -- when cards 
get swapped (even when he needs $\spadesuit$ to win). When the probability 
of a favorable outcome is the same before and after the swap then the states 
are equivalent (``$\approx$'') from his point of view. This leads to {\it subjective} 
 probabilities given by the ratio of the number of favorable outcomes to the total; 
$p_{\spadesuit}={1 \over 2}$. Laplace's approach based on observers ignorance and 
not on the actual physical state is often regarded as the sole justification of Bayesian view.
It is controversial. In particular, it does not reflect the actual `physical' state of the
system: As seen in Fig. 1b, states before and after the swap 
are not equivalent, ``$\neq$''. Fig 1c shows how quantum theory leads to probabilities based 
on the physical state when the system of interest when ${\cal S}$ 
is entangled with `the environment' ${\cal E}$. Such entanglement
would occur as a result of decoherence. When ${\cal S}$
and ${\cal E}$ are maximally entangled, the swap on ${\cal S}$ 
has no effect  on its state. This is clear, since its effect  can be undone without acting on ${\cal S}$ 
-- by a `counterswap' that involves only ${\cal E}$. The final states are 
simply the same, so the probability of the swapped states of ${\cal S}$ must also be the
same!  Envariance can be used to prove (see Theorem 2) that for such  `even'  
entangled states that have the same absolute values of the coefficients
(which makes them envariantly swappable), probabilities of mutually exclusive
alternatives (orthonormal Schmidt states) are equal. Generalization of 
to the case of unequal coefficients is straightforward and establishes 
Born's rule $p_k=|\psi_k|^2$.  }
\label{CardsFigure} 
\end{figure*}

\subsection{Envariance and the  `principle of indifference' }

The idea that invariance under swaps implies ignorance and hence probabilities
is old, and goes back at least to Laplace [17]. We illustrate it in Fig. 1a. The appeal to
invariance under swaps leads to a definition that is known as `standard' 
(or, sometimes, `classical' -- the adjective we reserve in this paper
for its other obvious meaning). In classical physics standard definition has
to be applied with apologies, as it refers to observers' {\it subjective lack
of knowledge} about the system, rather than to the objective properties of
state of the system {\it per se}. That the observer (player) does not care about 
(is indifferent to) swapping of the cards in a shuffled deck is the consequence 
of the fact that he does not know their  `face values'.  It is a subjective reason
-- not reflected in the objective symmetries of the actual state of the deck 
(see Fig. 1b). To recognise this, it is said that certain possible states are,
in his subjective judgement, `equally likely'.

As always in physics, subjectivity is a source of trouble. In case of the above
`principle of indifference' this trouble is exacerbated by the fact that the objective
classical state (e.g., of the deck of cards) is well defined and does not 
respect `symmetries of the state of mind' of the observer.

Subjectivity was the principal reason this  `standard definition of probabilities' 
has fallen out of
favor, and was largely replaced by the `relative frequency approach' [18-20].
We shall not discuss it in detail as yet, but we remind the reader that in this
approach probability of a certain event -- i.e., a certain outcome -- is given
by its relative frequency in an {\it infinite} ensemble.

There are problems with this strategy as well. My discomfort with relative
frequencies stems from the fact that infinite ensembles generally do not exist,
and, hence, have to be imagined -- i.e., must be {\it subjectively} extrapolated 
from {\it finite} sets of data. So, subjectivity cannot be convincingly
exorcised in this manner.

I bring up relative frequencies here only to note
that the strategy of using swaps to define equiprobable events would work also
in this setting. When a swap (i.e., re-labeling all $s_k$ as $s_l$ -- e.g, renaming
``heads'' as ``tails''
and {\it vice versa}) applied to an ensemble leaves all the relative
frequencies of the swapped states unchanged, their probabilities must have
been equal. In a sense, this observation may be regarded as an independent
motivation of the assumption (a) above.

In quantum physics, exact symmetry of a composite state -- envariance --
can be used to demonstrate that the observer need not care about swapping,
as envariant swapping provably cannot alter anything in his part of the
bipartite system when $\psi_{\cal SE}$ is `even' -- that is, of the form
given by Eq. (5). We illustrate it in Fig. 1c. As we have already seen, 
envariance makes ignorance and probabilities easier to define.  
The circularity of the `standard definition' is easier to circumvent
in the quantum setting. It is however important to understand what fixes
this `menu of options', and to find out when they can be regarded as
effectively classical. 

\subsection{Quantum and classical ignorance}

Motivating the need for probabilities using ignorance about a preexisting state
is often regarded as synonymous with the {\it ignorance interpretation}.  
We have relied on a similar approach. However, for our purpose
a narrowly classical definition of ignorance through an appeal to definite
classical possibilities one of which actually exists independently of observer
and can be discovered by his measurements without being perturbed is simply
too restrictive: Observer can be ignorant of the outcome of his future
measurement also when the system in question is quantum and when there
are no fixed pre-existing alternatives. He can then choose between various
non-commuting observables and sets of alternative `events' defined by
their complementary eigenstates. That choice of what to measure
determines what is it observer is ignorant about -- what sort of information 
his about-to-be-performed measurement will reveal.

Quantum definition of probabilities based on envariance is superior to the
above mentioned classical approaches because it justifies ignorance 
objectively, without appealing to observers' subjective `lack of knowledge'. 
Entangled quantum state of ${\cal SE}$ (or any pair of entangled systems)
can be perfectly known to the observer beforehand. There can be multiple 
records of that state spread throughout the environment, making it 
`operationally objective' -- simultaneously accessible to many observers. 
They can use its objective global properties to demonstrate -- employing real 
swaps one can carry out in the laboratory -- that the outcomes of some of 
the measurements one can perform are provably swappable, and, hence, 
equiprobable. In this sense, observers can directly {\it measure} probabilities 
of various outcomes without having to find out first what these outcomes are.

Note that this can be accomplished without an appeal to an ``ensemble'',
``likelihood'',  or any other surrogate for probability: All that is needed is a
single system $\cal S$, as well as appropriate ${\cal E}$ and ${\cal C}$.
The entangled states we have studied in the preceding section can be then 
created, and their symmetries can be verified through manipulations and 
measurements. Moreover, the measurements involved have outcomes that 
can be predicted with certainty. This is how the concept of probability enters 
our discussion. Observer is certain of the global state, and (using envariance)
can count the number of envariantly swappable (and, hence, equiprobable)
outcomes.
 
The aim of the observer is to use records of the outcomes of past measurements
(his data) that in effect define the global state to predict future events -- his future 
record of the measurement of the local system. Tools that can be
legally employed in this task include observers knowledge of the `no collapse'
quantum physics, encapsulated in the opening paragraph of Section II, as well as
the {\bf facts 1-3} of the preceding section. They do not include Born's rule:
Hence the trace operation, reduced density matrices, etc. are `off limits' in our
derivations (although when we succeed, embargo on their use will be lifted).

In the no-collapse settings, effective classicality of the memory can be
justified through appeal to decoherence [3-8,15]. But here we cannot appeal to
full-fledged decoherence, cannot rely on trace, etc. Is there still a way to
recover enough of the `effectively classical' to justify existence of
classical records we took for granted in the preceding section?

\subsection{Problems of `no collapse' approaches}

There were several attempts to make sense of probabilities in the `no collapse'
setting [21-25]. They have relied almost exclusively on relative frequencies
(counting probabilities) [21-23]. The aim was to show that, in the limit of
infinitely many measurements, only branches in which relative frequency 
definition would have given the answer for probabilities consistent with Born's rule 
have a non-vanishing measure. However, that meant dismissing infinitely many 
(`maverick') branches where this is {\it not} the case, because their amplitude
becomes negligible in the same infinite limit. All of these attempts have been
shown to use circular arguments [25-29]: In effect, they were all forced to
assume that relative weights of the branches are based on their amplitudes.
That meant that another measure was introduced, without physical justification,
in order to legitimize use of relative frequency measure based on counting.

By contrast, derivations of Born's rule that assume `collapse' either
explicitly [30,32] or implicitly [31], that is by considering {\it ab initio} infinite 
ensembles of identical systems so that `branches' with `wrong' relative
frequency of counts simply disappear as their amplitudes vanish in 
the infinite ensemble limit have been successful (although see [33]
for a somewhat more critical assessment of Refs. [31,32]). 

Indeed, Gleason's theorem [30] is now an accepted and rightly famous 
part of quantum foundations. It is rigorous -- it is after all a theorem about 
measures on Hilbert spaces. However, regarded as a result in physics 
it is deeply unsatisfying: it provides no insight into physical significance
of quantum probabilities -- it is not clear why the observer should assign 
probabilities in accord with the measure indicated by Gleason's 
approach [22-29,31-34]. This has motivated various primarily frequentist 
approaches [22-29,,31,32,34]. However, as was already noted in 
the discussion of the relative frequency approach, appeal to infinite
ensembles is highly suspect, especially when -- as is the case here -- the
desired effect is achieved only when the size of ensemble $N = \infty$ [27-29,33].

In view of these difficulties, some have even expressed doubts as to whether
there is any room for the concept of probability in the no-collapse setting
[26-29]. This concern is on occasion traced all the way to the {\it identity}
of the observer who could define and make use of probabilities in Many Worlds
universe. Such criticism was often amplified by pointing out that,
in the pre-decoherence versions of Everett - inspired interpretations,
decomposition of the universal state vector is not unique (see e.g. comments
in Ref. [27]) so it is not even clear what `events' should such probabilities
refer to. This difficulty was regarded by some as so severe
that even staunch supporters of Everett considered adding an {\it ad hoc} rule
to quantum theory to specify a preferred basis (see e.g. Ref. [35]) in clear
violation of the spirit of the original relative state proposal [12].

Basis ambiguity was settled by einselection [3-9,14,15]. But standard tools of
decoherence are `off limits' here. And even if we were to accept decoherence -
based resolution of basis ambiguity and used einselected pointer states to
define branches, we would be still faced with at least one remaining aspect of
the `identity crisis': In course of a measurement the memory of the observer
`branches out' so that it appears to record (and, hence, critics conclude,
observer should presumably perceive) {\it all} of the outcomes. If that were so,
then there would be no room for distinct outcomes, choice, and hence, no need
for probabilities.

{\it Existential interpretation} is based on einselection [3,8,15] and can
settle also this aspect of the identity crisis. It follows physical state
of the observer by tracking its evolution, also in course of measurements.
Complete description of observers' state includes physical state of his memory,
which is updated as he acquires new records. So, {\it what the observer
knows is inseparable from what the observer is}. Observer who has acquired new
data does not lose his identity:  He simply extends his records -- his
history. Identity of the observer as well as his ability to measure persist
as long as such updates do not result in too drastic a change of that state
(see Wallace [36] for a related point of view).  For instance, Wigner's friend
would be able to continue to act as observer with reliable memory in each
`outcome branch', but the same cannot be said about the Schr\"odinger's cat.

But the existential interpretation depends on decoherence. Threfore, we have to
either disown it altogether while embarking on the derivation of Born's rule, 
or re-establish its foundations without relying on trace and reduced density 
matrices to derive Born's rule. Once we are successfull, we can then proceed
to reconstruct all of the ``standard lore" of decoherence, including all of its 
interpretational implications, on a firm and deep foundation -- quantum
symmetry of entangled states.

We note that when (at the risk of circularity we have detailed above)
decoherence is assumed, Born's rule can be readily derived using variety
of approaches [8,24] including versions of both standard and frequentist
points of view [8] as well as approaches based on decision theory [24,36]
suggested by Deutsch and Wallace. I shall comment on these in Section VII.

\section{Pointer states, records, and `events'}

Probability is a tool observers employ in the absence of certainty to predict
their future (including in particular future state of their own memory) using
the already available data -- present state of their memory. Sometimes these
data can determine some of the future events uniquely. However, often
predictions will be probabilistic -- the observer may count on one of
the potential outcomes, but will not know which one. Our task is to replace
questions that uncertain outcomes of future measurement on the system
with situations that allow for a certainty of prediction about the effect of
some actions (swaps) on an enlarged system. In that sense we are reducing 
questions about probabilities in general to the straightforward case -- to the
questions that yield answers with certainty (i.e., with {\it probability} of 1). This special
case makes the connection with probability in the situation when probability 
is known. It provides us with the overall normalization. Using this connection,
we then infer probabilities of possible outcomes of measurements on ${\cal S}$ 
from the analogue of the Laplacean `ratio of favorable events to the total number 
of equiprobable events', which we shall see in Section V is a good definition of 
quantum probabilities for events associated with effectively classical records 
kept in pointer states. In quantum physics this ratio can be determined using 
envariance and even verified experimentally prior to finding out what the event is. 
But we do need events. Hence, we need pointer states.

In classical physics it can be always assumed that a random event was
pre-determined -- i.e., that the to-be-discovered state existed objectively
beforehand and was only revealed by the measurement. Indeed, such objective
preexistence is often regarded as a pre-condition of the `ignorance
interpretation' of probabilities. In quantum physics, as John Archibald Wheeler
put it paraphrasing Niels Bohr, ``No phenomenon is a phenomenon untill it is
a recorded phenomenon" [37]. Objective existence cannot be attributed 
to quantum states of isolated, individual systems. When the system is being 
monitored by the environment, one may approximate objective existence 
by selectively proliferating quantum information (this is the essence of 
{\it quantum Darwinism}). Operational definition of objectivity -- one of the key
symptoms of classicality -- can be recovered using this idea, as selective
proliferation leads to many copies of the information about the same
`fittest' observable.

Here we deal with individual quantum events -- outcomes of future
measurements on individual quantum systems, not carried out as yet. They
need not (and, in general, do not) preexist even in the sense of quantum
Darwinism -- selective proliferation of information has not happened as yet.
The observer will decide what observable he will measure (and, hence, what
information will get proliferated, and become effectively classical). The menu of
events should be determined by the observable he chooses to measure.

Objective pre-existence of events is not needed {\it before} the measurement.
On the other hand, as noted above, we need some aspects of classicality to
represent memory of the observer {\it after} the measurement. Objectivity
would be best, but we can settle for less: The key ingredient is the
existence of well-defined `events' -- record states that, following measurement,
will reliably preserve correlation with the recorded state of the system
in spite of the immersion of the memory in the environment. Pointer 
states [5-9] are called for, but we are not allowed to use decoherence.

There are at least three reasons why preferred states (that are stable in spite
of the opennes of the system) are essential. The original reason for their
introduction [5] was  to assure that observers' memory is effectively classical 
-- einselection of preferred basis ascertains that his `hardware' 
can keep records only in the pointer basis.

It follows that model observers will store and process information more or less
like us (and more or less like a classical computer). For instance, human brain
is a massivelly parallel, and as yet far from understood, but nevertheless
classical information processing device. And classical records cannot exist
(cannot be accessed) in superpositions. 

A more pertinent reason to look for
preferred states is the recognition that the information processing hardware
of observers is open -- immersed in the environment. Interaction with 
${\cal E}$
is a fact of life. Unless we find in the memory Hilbert space `quiet corners'
that remain quiet in spite of this openness, reliable memory (and 
hence reliable
information processing) will not be possible. 

Last not least, the very
idea of measuring makes sense only if measurement outcomes can be used
for prediction. But most of the systems of interest are `open' as well.
Therefore, predictability can be hoped for only in special cases -- for the
einselected pointer states of systems that can remain correlated with
the apparatus that has measured them [5,6,8].

Decoherence methods used to analyse consequences of such immersion
in the environment employ reduced density matrices and trace [3-9,14,15].
They are based on Born's rule. Once decoherence is assumed, Born's rule can be
readily derived [8], but this strategy courts circularity [3,38]. Obviously,
if we temporarily renounce decoherence (so that we can attempt to derive
$p_k = |\psi_k|^2$) we have to find some other way to either justify existence
of pointer states and the einselection - based definition of branches in
a decoherence - independent manner, or give up and conclude that while Born's
rule is consistent with decoherence, it cannot be established from more
basic principles in a no-collapse setting. We shall show that, fortunately,
preferred states do obtain more or less directly from envariance. That is,
environment - induced superselection, and, hence, pointer basis and branches
can be defined by analysing correlations between quantum systems, very much
in the spirit of the original `pointer basis' proposal [5], and without
the danger of circularity that may have arisen from reliance on trace
and reduced density matrices.

\subsection{Relative states and definite outcomes: Modeling the collapse}

The essence of the `no collapse' approach is its {\it relative state} nature --
its focus on correlations. Once we identify possible record
states of the memory, the correlation between them and the states of the
relevant fragments of the Universe is all that matters: Observer will be
`aware' of what he knows, and all he knows is represented by his data.
The key is, however to identify relative {\it pointer states} that are
suitable as memory states.

We start our discussion with a confirmatory pre-measurement in which there
is no need for the full-fledged `collapse'. Observer is presented with a system
in a state $\varphi$, and told to measure observable that has $\varphi$ as one
of its eigenstates. He need not know the outcome beforehand. Given a set of
potential mutually exclusive outcomes $\{\dots,\varphi,\dots \}$ observer
can devise an interaction that will -- with certainty -- lead to:
$$ |A_0\rangle^{\otimes {\cal N}} |\varphi\rangle  \
{\buildrel {\{\dots\varphi\dots\}} \over {\longrightarrow }}
|A_0\rangle^{\otimes {\cal N}-1} |A_{\varphi}\rangle
|\varphi\rangle \ . \eqno(11a)$$
Above, `ready to record' memory cells are designated by $|A_0\rangle$, and the
symbol above the arrow representing measurement indicates that the set of
possible outcomes includes $\varphi$. This pre-measurement can be repeated
many ($m$) times:
$$ |A_0\rangle^{\otimes {\cal N}} |\varphi\rangle
\ \ {\buildrel {\{\dots \varphi \dots\}} \over
{\underbrace {\longrightarrow \dots \longrightarrow}_m}} \ \
|A_0\rangle^{\otimes {\cal N}-m} |A_{\varphi}\rangle^{\otimes m}
|\varphi\rangle \ . \eqno(11b)$$
Each new outcome can be predicted by the observer from the first record 
providing that -- as we assume here and below -- there is no evolution 
between measurements and that the measured observable does not change.

In spite of the concerns we have reported before there is no threat to
the identity of the observer. Moreover, in view of the recorded evidence,  
Eq. (11b), and his ability to predict future outcomes in this situation,
observer may assert that {\it the system is in the state} $\varphi$.
Indeed, if we did not assume that our observer is familiar with 
quantum physics,
we could try and convince him that -- in view of the evidence -- the system
must have been in the preexisting `objective state' $\varphi$ already before
the first measurement. This happens to be the case in our example, but the
state of a single system is not `objective' in the same absolute sense one may
take for granted in the classical realm: Obviously, in quantum theory a long
sequence of records after the measurement is no guarantee of  the `objective 
preexistence' of the observer - independent state before the measurement
in the same sense as was the case in the classical realm.

When the observer subsequently decides to measure a different observable with
the eigenstates $\{|s_k\rangle\}$ such that:
$$ |\varphi\rangle  = \sum_{k=1}^N \langle s_k|\varphi\rangle
|s_k\rangle \eqno(12a)$$
the overall state viewed `from the outside' will become:
$$ |A_0\rangle^{\otimes {\cal N}-m} |A_{\varphi}\rangle^{\otimes m}
|\varphi\rangle $$
$$ {\buildrel {\{|s_k\rangle\}} \over {\longrightarrow }} 
|A_0\rangle^{\otimes {\cal N}-m-1} |A_{\varphi}\rangle^{\otimes m}
\sum_{k=1}^N \langle s_k|\varphi\rangle |A_k\rangle
|s_k\rangle \ . \eqno(13a)$$
It is tempting to regard each sequence of records as a `branch' of the above
state vector. It certainly represents a possible state of the observer with
the data concerning his `history', including records of the state of 
the system. From his point of view, observer can therefore describe this 
`personal' recorded history
as follows: `As long as I was measuring observables with the eigenstate
$\varphi$, the outcome was certain -- it was always the same. After the first
measurement, there was no surprise. In that operational sense the system
{\it was} in the state $|\varphi\rangle$. When I switched to measuring
$\{|s_k\rangle\}$, the outcome became (say) $|s_{17}\rangle$, and the system
was in that state thereafter.' This conclusion follows from his records:
$$ \dots |A_{\varphi}\rangle^{\otimes m} |\varphi\rangle  \
{{\buildrel {\{|s_k\rangle\}} \over  {\longrightarrow}}}
\dots |A_{\varphi}\rangle^{\otimes m} |A_{17}\rangle |s_{17}\rangle$$
$${{\buildrel {\{|s_k\rangle\}} \over
{\underbrace {\longrightarrow \dots \longrightarrow}_{l-1}}}} \ \
\dots |A_{\varphi}\rangle^{\otimes m}
|A_{17}\rangle^{\otimes l} |s_{17}\rangle \ . \eqno(13b)$$
There will be, of course, $N$ such branches, each of them labeled by a specific
record (e.g., $\dots |A_{16}\rangle^{\otimes l} ,~ |A_{17}\rangle^{\otimes l},~
|A_{18}\rangle^{\otimes l} \dots $). But, especially when records
are well - defined and stable, then there is again no threat to the identity
of the observer. (Also note that in the equation above we have stopped
counting the ``still available'' empty memory cells. In the tradition that
dates back to Turing we shall assume, from now on, that observer has
enough  `blank memory' to record whatever needs to be recorded.)

Now come the two key lessons of this section: To `first approximation', from
the point of view of the evidence (i.e., records in observer possession) 
there is no difference between the `classical collapse' of many future 
possibilities into one present actuality he experiences after his first 
`confirmatory' measurement, Eq. (11), and the `quantum collapse', Eq. (13). 
Before the confirmatory measurement observer was told what observable 
he should measure, and he found out the state $\varphi$. A very similar
thing happened when the observer switched to the different observable with
the eigenstates $\{|s_k\rangle\}$: After the first measurement with
an uncertain outcome a predictable sequence of confirmations followed.

In the `second approximation', there is however an essential difference: In
case of the classical collapse, Eq. (11), the observer had {\it less} of an idea
about what to expect than in the case of genuine quantum measurement, Eq. (13).
This is because now -- in contrast to the ``ersatz collapse'' of Eq. (11) --
observer knows all that can be known about the system he is measuring. Thus,
in case of Eq. (11), observer did not know what to expect: He had no certifiably
accurate information that would have allowed him to gauge what will happen. He
could have been left completely in the dark by the preparer (in which case he
might have been swayed by the arguments of Laplace [17] and to assign equal
probabilities to all conceivable outcomes), or might have been persuaded to
accept some other Bayesian priors, or could have been even deliberately misled.
Thus, in the classical case of Eq. (11) the observer has no reliable source of
information -- no way of knowing what he actually does not know -- and, hence,
no {\it a priori} way of assigning probabilities. By contrast, in the quantum
case of Eq. (13) observer knows all that can be known. Therefore, he knows as
much as can be known about the system -- he is in an excellent position to
deduce the chances of different conceivable outcomes of the new measurement
he is about to perform.

This is the second lesson of our discussion, and an important conclusion:
In quantum physics, ignorance can be quantified more reliably than
in the classical realm. Remarkably, while the criticisms about assigning
probabilities `on the basis of ignorance' have been made before
in the classical context of Eq. (11), perfect knowledge available in the
quantum case was often regarded as an impediment.

\subsection{Environment - induced superselection without decoherence}

We shall return to the discussion of probabilities shortly. However, first we
need to deal with the preferred basis problem pointed out above. The 
prediction observer is trying to make -- the only prediction he can 
ultimately hope to
verify -- concerns the future state of his memory ${\cal A}$. We need to
identify -- without appeals to full-fledged decoherence -- which memory states
are suitable for keeping records, and, more generally, which states in the rest
of the Universe are sufficiently stable to be worth recording. The criterion:
such states should be sufficiently well behaved so that the correlation between
observers records and the measured systems should have predictive power [5].
Predictability shall remain our key concern (as was the case in the studies
employing decoherence [3-9,14,15,39]) although now we shall have to formulate
a somewhat different -- `decoherence - free' -- approach.

The issue we obviously need to address is the basis ambiguity: Why is it more
reasonable to consider a certain set $\{|A_k\rangle\}$ as memory states rather than
some other set $\{|B_k\rangle\}$ that spans the same (memory cell or pointer
state) Hilbert space? One answer
is that observer presumably tested his memory before, so that the initial state
of his record - bearing memory cells as well as the interaction Hamiltonian
used to measure -- to generate conditional dynamics -- have the structure that
implements the `truth table' of the form:
$$ |A_0\rangle |s_k\rangle \longrightarrow |A_k\rangle |s_k\rangle \ .
\eqno(14) $$
But this is not a very convincing or fundamental resolution of the broader
problem of the emergence of the preferred set of effectively classical
(predictable) states from within the Hilbert space. Clearly, the state
on the RHS of Eq. (13a) is entangled. So one could consult ${\cal A}$
{\it in any basis} $\{|B_l\rangle\}$ and discover the corresponding state of
${\cal S}$:
$$|r_l\rangle=\sum_{k=1}^N \langle B_l|A_k\rangle  \langle s_k|\varphi\rangle
|s_k\rangle \ . \eqno(15)$$
The question is: Why should the observer remember his past, think of his future
and perceive his  present state in terms of $\{ |A_k \rangle \}$'s rather than
$\{|B_l\rangle\}$'s? Future measurements of one of the states $\{|r_l\rangle\}$
(which form in general a non-orthogonal, but typically complete set within the
subspace spanned by $\{|s_k\rangle\}$) could certainly be devised so that
the outcome confirms the initial result. Why then did we write down the chain
of equations (13b) using one specific set of states (and anticipating
the corresponding set of effectively classical correlations)?

\subsubsection{Pointer states from envariance: The case of perfect correlation}

Basis ambiguity is usually settled by an appeal to decoherence. That is, after
the pre-measurement -- after the memory ${\cal A}$ of the apparatus or of the
observer becomes entangled with the system -- ${\cal A}$ interacts with the
environment ${\cal E}$, so that ${\cal E}$ in effect pre-measures ${\cal A}$:
$$ |A_0\rangle|\varphi\rangle |{\varepsilon}_0\rangle\
{\buildrel {\{|s_k\rangle\}} \over {\longrightarrow }} \
\bigl(\sum_{k=1}^N \langle s_k|\varphi\rangle |A_{k}\rangle|s_k\rangle \bigr)
|{\varepsilon}_0\rangle $$
$$ {\buildrel {\{|\varepsilon_k\rangle\}} \over {\longrightarrow }} 
\sum_{k=1}^N \langle s_k|\varphi\rangle |A_{k}\rangle
|s_k\rangle |{\varepsilon}_k \rangle\ = |\Psi_{\cal SAE}\rangle \ . \eqno(16)$$
The observable left unperturbed by the interaction with the environment
should be the `record observable' of ${\cal A}$. Then the preferred pointer
states $\{|A_k\rangle\}$ are einselected.

Note that this very same combined state would have resulted if the environment
interacted with ${\cal S}$ and `measured it' directly in the basis
$\{|s_k\rangle\}$, before the correlation of ${\cal A}$ and ${\cal S}$.
Clearly, there is more than one way to ``skin the Schr\"odinger cat".
This remark is meant to motivate and justify a simplification of notation
later on in the discussion. In reality, it is likely that both ${\cal A}$
and ${\cal S}$ would have interacted with their environments, and that each
might be immersed in (one or more) physically distinct ${\cal E}$'s.
Recognising this in notation is cumbersome, and has no bearing on our immediate
goal of showing that einselection of pointer states can be justified without
taking a trace to compute the reduced density matrix.

In the situation when the measured observable is hermitean
(so that its eigenstates are orthogonal) and environment pre-measures
the memory in the record states, Eq. (16) has the most general form possible.
It is therefore enough to focus on a single ${\cal E}$, and reserve the right
to entangle it sometimes with ${\cal A}$ and sometimes with ${\cal S}$, and
sometimes with both.

Given that the measured observable is hermitean and recognising the nature
of the conditional dynamics of the ${\cal AE}$ interaction, the resulting
density matrix would have the form:
$$ \rho_{\cal SA}=Tr_{\cal E}|\Psi_{\cal SAE}\rangle \langle\Psi_{\cal SAE}|
  = \sum_{k=1}^N |\langle s_k|\varphi\rangle|^2
~|A_{k}\rangle  |s_k\rangle \langle s_{k}| \langle A_{k}| \ . \eqno(17)$$
Preservation of the perfect correlation between ${\cal S}$ and ${\cal A}$ in
the preferred set of pointer states of ${\cal A}$ is a hallmark of a 
successfull
measurement. Entanglement is eliminated by decoherence, (e.g. discord 
disappears
[41]) but one-to-one classical correlation with the einselected states remains.
This singles out $\{|A_k\rangle\}$'s as `buds' of the new branches that can be
predictably extended by subsequent measurements of the same observable, e.g.;
$$ \dots |A_{17}\rangle |s_{17}\rangle |\varepsilon_{17}\rangle
{{\buildrel {\{|s_k\rangle\}} \over
{\underbrace {\longrightarrow \dots \longrightarrow}_l}}}
\dots |A_{17}\rangle^{\otimes l}
|s_{17}\rangle |\varepsilon_{17}\rangle^{\otimes l} \ . $$
The trace in Eq. (17) that yields density matrix is however `illegal' in
the discussion aimed at the derivation of Born's rule: Both the physical
interpretation of the trace and of the reduced density matrix are justified
assuming Born's rule. So we need to look for a different, more fundamental
justification of the preferred set of pointer states.

Fortunately, envariance alone hints at the existence of the preferred states:
As Theorem 1 demonstrates, phases of Schmidt coefficients have no relevance
for the states of entangled systems. In that sense, superpositions of Schmidt
states for the system ${\cal S}$ or for the apparatus ${\cal A}$ that are
entangled with some ${\cal E}$ do not exist either. Now, this corollary of
Theorem 1 comes close to answering the original pointer basis 
question [5]: When
measurement happens, into what basis does the wavepacket collapse? Obviously,
if we disqualify all superpositions of some preferred set of states, only these
preferred states will remain viable. This is an envariant version of 
the account of the negative selection process, the {\it predictability sieve} 
that is usually introduced using full-fledged decoherence. 
The aim of this section is to refine this
envariant view of the emergence of preferred basis [2,3] by investigating
stability and predictive utility of the correlations between candidate record
states of the apparatus and the corresponding state of the system.

The ability to infer the state of the system from the state of the apparatus is
now very much dependent on the selection of the measurement of ${\cal 
A}$. Thus,
the {\it preferred basis $\{|A_k\rangle\}$ yields one-to-one correlations between
${\cal A}$ and ${\cal S}$ that do not depend on ${\cal E}$};
$$ |A_k\rangle \langle A_k| |\Psi_{\cal SAE}\rangle \propto \
|A_k\rangle |s_k\rangle |\varepsilon_k \rangle \ . \eqno(18a)$$
Above, we have written out explicitely just the relevant part of the resulting
state -- i.e., the part that is selected by the projection.

By contrast, if
we were to rely on any other basis, the information obtained would have to do
not just with the system, but with a joint state of the system and
the environment. For instance, measurement in the complementary basis:
$$ |B_l\rangle  =  \sum_{k=1}^N {{e^{i 2\pi kl/N} }  \over {\sqrt N}}  |A_k\rangle
$$
would result in rather messy entangled state of ${\cal S}$ and ${\cal E}$:
$$ |B_l \rangle \langle B_l | |\Psi_{\cal SAE}\rangle \propto \ |B_l\rangle
\bigr( \sum_{k=1}^N {{e^{-i 2\pi kl/N}}   \over {\sqrt N}} 
a_k|s_k\rangle|\varepsilon_k\rangle \bigl) \ . \eqno(18b)$$
This correlation with entangled ${\cal SE}$ state obviously prevents one
from inferring the state of the system of interest ${\cal S}$ alone from
memory states in the basis $\{|B_l\rangle\}$: {\it Bases other than pointer
states $\{|A_k\rangle \}$ do not correlate solely with ${\cal S}$, but
with entangled (or, more generally, mixed but correlated) states of 
${\cal SE}$}.

We conclude that the pointer states of ${\cal A}$ can be defined in 
the ``old fashioned way'' -- as states best at preserving correlation 
with ${\cal S}$, and nothing but ${\cal S}$. This perfect correlation will persist
even when, say, the environment is initially in a mixed state. Our decoherence
- free definition of preferred states appeals 
to the same intuition as the original argument in [5], or as 
the predictability sieve [15,39,3,8,9], but does not require reduced 
density matrices or other ingredients that rely on Born's rule.

\subsubsection{Pointer states from envariance: Einselection, records, and dynamics}

The example of einselection considered above allowed us to recover pointer
states under very strong assumptions, i.e., when
$\{|s_k\rangle\},~\{|A_k\rangle\}$ and $\{|\varepsilon_k\rangle\}$ are all
orthonormal. But such tripartite Schmidt decompositions are an exception [40].
It is therefore important to investigate whether our conclusions remain
valid when we look at a more realistic case
of an apparatus that first entangles with ${\cal S}$ through pre-measurement
and thereafter continues to interact with ${\cal E}$. The ${\cal SA}$
interaction may be brief, and can be represented by Hamiltonians that induce
conditional dynamics summed up in the truth table of Eq. (14). The effect of
immersion of the apparatus pointer or the memory cell in the environment is
often modelled by an on-going interaction generated by the Hamiltonian with
the structure:
$$ H_{\cal AE} = \sum_{k, \nu} g_{k \nu} |A_k\rangle \langle A_k| \otimes
|e_{\nu}\rangle \langle e_{\nu}| \ \eqno(19a)$$
which leads to a time-dependent state:
$$ |\Psi_{\cal SAE}\rangle = \sum_k a_k |s_k\rangle |A_k \rangle \sum_{\nu}
\gamma_{\nu} e^{- i g_{k\nu} t} |e_{\nu}\rangle $$
$$ = \sum_k a_k |s_k\rangle |A_k\rangle |\varepsilon_k(t)\rangle \ . 
\eqno(19b)$$
The {\it decoherence factor}:
$$ \zeta_{kk'} = \langle \varepsilon_k | \varepsilon_{k'}\rangle =
\sum_{\nu} |\gamma_{\nu}|^2  e^{i(g_{k'\nu} - g_{k\nu})t} \eqno (19c)$$
responsible for suppressing off-diagonal terms in the reduced density matrix
$\rho_{\cal SA}$ is generally non-vanishing although, for sufficiently large
$t$ and large environments, it is typically vanishingly small. Of course,
as yet -- i.e., in the absence of Born's rule -- we have no right
to attribute any physical significance to its value.

In the above tripartite state only $\{|A_k\rangle\}$ are by assumption
orthonormal. We need not assert that about $\{|s_k\rangle\}$ -- the initial
truth table might have been, after all, imperfect, or the measured observable
may not be hermitean. Still, even in this case
with relaxed assumptions perfect correlation of the states of the system with
the fixed set of records -- pointer states $\{|A_k\rangle\}$ -- persists, as
the reader can verify by repeating calculations of Eq. (18).

The decomposition of Eq. (19b) is typically not Schmidt when we regard
${\cal SAE}$ as bipartite, consisting of ${\cal SA}$ and ${\cal E}$ -- product
states $|A_k\rangle |s_k\rangle$ are orthonormal, but the scalar 
products of the
associated states $|\varepsilon_k(t)\rangle$ are given by the 
decoherence factor
$\zeta_{kk'}(t)$, which is generally different from zero. The key 
conclusion can
be now summed up with:

\noindent{\bf Theorem 3:} When the environment-apparatus interaction 
has the form
of Eq. (19a), only the pointer observable of ${\cal A}$ can maintain perfect
correlation with the states of the measured system independently of the initial
state of ${\cal E}$ and time $t$.

\noindent{\bf Proof:} From Eq. (19b) it is clear that states $|A_k\rangle$
maintain perfect 1-1 correlation with the states $|s_k\rangle$ at all times,
and for all initial states of ${\cal E}$ -- i.e., independetly of
the coefficients $\gamma_k$. This establishes that $|A_k\rangle$ are good
pointer states.

To complete the proof, we still need to establish the converse, i.e., that
these are the {\it only} such good record states. To this end, consider another
set of candidate memory states of ${\cal A}$:
$$ |B_l\rangle = \sum_k b_{lk} |A_k\rangle \ . $$
The corresponding conditional state of the rest of ${\cal SAE}$ is of the form:
$$ |B_l\rangle \langle B_l | \Psi_{\cal SAE} \rangle \propto |B_l\rangle \sum_k a_kb_{lk} |s_k\rangle
|\varepsilon_k(t)\rangle \ . \eqno(19d)$$
It is clearly an entangled state of ${\cal SE}$ unless all $k,~k'$ that appear
with non-zero coefficients differ only by a phase. That condition implies that
$\langle \varepsilon_k (t) | \varepsilon_{k'}(t)\rangle = \exp (i \phi_{kk'})$,
that is, $|\zeta_{kk'}(t)|=1$. This would in turn mean that the environment
has not become correlated with the states $|A_k\rangle$.

In general $|\zeta_{kk'}(t)|<1$. Thus, states of ${\cal SE}$ correlated with
the basis $|B_l\rangle$ of the apparatus are entangled -- records contained in
that basis do not reveal information about the system alone (as records in the
basis $|A_k\rangle$ do), but, rather, about ill-defined entangled states
of the system of interest ${\cal S}$ and environment ${\cal E}$ (which is of
no interest). Thus, as stated in the thesis of this theorem, the only way to
assure 1-1 record - outcome correlation for all $t$ independently of 
the initial
state of ${\cal E}$ is to use pointer states $|A_k\rangle$. QED.

\noindent{\bf Remark:} The above argument sheds an interesting new light
on the nature of the environment - induced loss of information. 
In case of imperfections the state correlated with records kept in 
the memory has the form of Eq. (19d) -- that is, it can be a {\it pure}
entangled state of ${\cal SE}$, and not the state of ${\cal S}$ alone. 
Observer still knows exactly the state, but it is a state of a wrong
system! Instead of the ``system on interest'' ${\cal S}$, it is a composite
${\cal SE}$. Consequently, (i) even if the observer knew the initial 
state of ${\cal E}$, the state of the system of interest ${\cal S}$ cannot be 
immediately deduced. Moreover, (ii) typically the initial state of ${\cal E}$ 
is not known, which makes the state of ${\cal S}$ even more difficult 
to find out.

We have provided a definition of pointer states by examining the structure of
correlations in the pure states involving ${\cal S, A}$ and ${\cal E}$. It is
straightforward to extend this argument
to the more typical case when ${\cal E}$ is initially in a mixed state
(for example, by ``purifying'' ${\cal E}$ in the usual manner). So, all of our
discussion can be based on pure states and projections -- preferred pointer
states emerge without invoking trace or reduced density matrices.

In spite of the more basic approach, our motivation has remained the same:
Preservation of the classical (or at least ``one way classical'' [41])
correlations, i.e., correlations between the preferred orthonormal set of
pointer states of the apparatus and of the (possibly more general) states
of the system that can be accessed (by measuring pointer observable of
${\cal A}$) without further loss of information. Therefore, it is perhaps not
too surprising that the preferred pointer observable
$$ \Lambda = \sum_k \lambda_k |A_k\rangle \langle A_k| $$
commutes with the ${\cal AE}$ interaction Hamiltonian, Eq. (19a), responsible
for decoherence:
$$ [\Lambda, H_{\cal AE}] = 0 \ . \eqno(20) $$
This simple equation valid under special circumstances was derived in the paper
that has introduced the idea of pointer states in the context of quantum
measurements [5], starting from the same condition of the preservation of
correlations we have invoked here.

Pointer states coincide with Schmidt states in the tripartite ${\cal SAE}$
when a perfect premeasurement of a hermitean observable is followed by perfect
decoherence, that is when $\zeta_{kk'}=0$. This case was noted and its more 
general consequences were anticipated by brief comments about pointer states
emerging from envariance in [2,3].

We also note that the recorded states of ${\cal S}$ do not need to be
orthonormal for the above argument to go through. We have relied on the
orthogonality of the record states $\{|A_k\rangle\}$ and invoked some of
its consequences without having to assume idealized perfect measurements of
hermitean observables. We note that this strategy can be used to attribute
probabilities to non-orthogonal (i.e., overcomplete) basis states of
${\cal S}$, as long as they eventually correlate with orthonormal 
record states.

The ability to recover pointer states without appeal to trace -- and, hence,
without an implicit appeal to Born's rule -- is a pivotal consequence of
envariance: We have just shown that einselection can be deduced and `branches'
may emerge without relying on decoherence. Rather than use trace and reduced
density matrices we have produced a derivation based solely on the ability of
an open system (in our case memory ${\cal A}$) to maintain correlations with
the test system ${\cal S}$ in spite of the interaction with the environment
${\cal E}$.

In hindsight, this ability to find preferred basis without reduced density
matrices -- although unanticipated [29,38] -- can be readily understood:
{\it Emergence of preferred states} (which in the usual
decoherence - based approach habitually appear on the diagonal of the reduced
density matrix) {\it is a consequence of the disappearance of the off-diagonal
terms.} But off-diagonal terms disappear when states of the environment
correlated with them are orthonormal -- when the decoherence factor defined
by the scalar product $\langle\varepsilon_k|\varepsilon_l\rangle$ disappears
if $k\neq l$. So, as long as we do not use trace to assign weights
(probabilities) to the remaining (diagonal) terms, we are not invoking Born's
rule. In short, one can find out eigenstates of the reduced density matrix
without enquiring about its eigenvalues, and especially without regarding them
as probabilities.

\section{Probability of a future record}

To arrive at Born's rule within the `no-collapse' point of view we now follow
envariant strategy of Section II, but with one important difference: The probability
refers explicitely to the possible future state of the observer. This
approach is very much in the spirit of the {\it existential interpretation}
which in turn builds on the idea (that was most clearly stated by Everett
[12]) to let quantum formalism dictate its interpretation.

Existential interpretation is introduced more carefully and discussed
in detail elsewhere [15,8,3]. It combines relative state point of view with
the recognition of the emergence and role of the preferred pointer states.
In essence, and in the context of the present discussion, according to
the existential interpretation observer will perceive himself (or, to be more
precise, his memory) in one of the pointer states, and therefore attribute
to the rest of the Universe state consistent with his records. His memory
is immersed in the environment. His records (stored in pointer states) are 
not secret. Indeed, because of the persistent monitoring by the environment, 
many copies of the information inscribed in his possession are `in public
domain' -- in the environmental degrees of freedom: Hence, the content of his
memory can be in principle deduced (e.g., by many other observers, monitoring
independent fragments of the environment) from measurements of the relevant
fragments of the environment. Selective proliferation 
of information about pointer observables -- allows records to be in effect
`relatively objective' [3,8]. This operational notion of objectivity is all
that is needed for the `objective classical reality' to emerge. Obviously,
observer will not be able to redefine memory pointer states -- correlations
involving their superpositions are useless for the purpose of prediction.

Existential interpretation is usually justified by an appeal to decoherence,
which limits the set of states that can retain useful correlations to
pointer states -- states that can persist, and therefore, exist -- to a small
subset of all possible states in the memory Hilbert space. As we have seen
in the preceding section, the case for persistence of the pointer states can
be made by exploiting nature -- and, especially, persistence -- of the
${\cal SA}$ correlations established in course of the measurement in spite
of the subsequent interaction of with the environment. In short, we can ask
about the probability that the observer will end up in a certain pointer state
-- that his memory will contain the corresponding record, and the rest
of the Universe will be in a state consistent with these records -- without
using trace operation or reduced density matrices of full-fledged decoherence.

\subsection{The case of equal probabilities}

We begin with the case of equal probabilities. That is, we consider:
$$|\varphi\rangle \propto \sum_{k=1}^N  e^{i \phi_k} |s_k\rangle  \ ,
\eqno(12b)$$
and note that following pre-measurement and as a consequence of the resulting
entanglement with the environment we get:
$$|\bar\Psi_{\cal SAE}\rangle= \sum_{k=1}^N e^{i \phi_k}
|s_k\rangle|A_{k}\rangle |\varepsilon_k\rangle=\sum_{k=1}^N e^{i \phi_k}
|s_k,A_{k}(s_k)\rangle |\varepsilon_k\rangle\  . \eqno(21)$$
The notation on the RHS above is introduced temporarily to emphasize that the
essential unpredictability observer is dealing with concerns his future state.
As the record states $|A_k\rangle$ are orthogonal, explicit recognition of this
focus of attention allows one
to consider probabilities of more general measurements where the outcome states
$\{|s_k\rangle\}$ of the system are not necessarily associated with
the orthonormal eigenstates of hermitean operators.

Orthogonality of states associated with events is needed to appeal to
envariance -- we want events to be `mutually exclusive'. Now it can be provided
by the record states. Both measurements of non-hermitean observables as well as
``destructive measurements'' that do not leave the system in the eigenstate
of the measured observable belong to this more general category. For instance:
$$|\bar\varphi\rangle |A_0\rangle|\varepsilon_0\rangle \propto
\bigl(\sum_{k=1}^N  e^{i \phi_k} |s_k\rangle\bigr) 
|A_0\rangle|\varepsilon_0\rangle$$
$$\longrightarrow |s_0\rangle \sum_k 
|A_k(s_k)\rangle|\varepsilon_k\rangle \eqno(22) $$
is a possible idealized representation of a `destructive' measurement such
as a photodetection that leaves the relevant mode of the field in the vacuum
state $|s_0\rangle$ but allows the detector to record pre-measurement state
of ${\cal S}$.

Most of our discussion below will be applicable to such more general (and more
realistic) measurements. However, the most convincing `existential' evidence
for the `collapse' of the state of the system is provided by non-demolition
measurements. In that case repeating the same measurement will yield the same
outcome:
$$|\bar\Psi_{{\cal SA}^{\otimes l}{\cal E}}\rangle\ \
=\ \sum_{k=1}^N  e^{i \phi_k}
|s_k\rangle|A_{k}\rangle^{\otimes l} |\varepsilon_k\rangle^{\otimes l}
\ . \eqno(23)$$
The observer will be led to conclude on the basis of his record, (e.g.
$\dots |A_{\varphi}\rangle^{\otimes m} |A_{17}\rangle^{\otimes l}$, Eq. (13b))
that a `collapse' of the state from the superposition of the potential outcomes
(Eq. (12b)) into a specific actual outcome ($|s_{17}\rangle$) has ocurred.
From his point of view, this clearly is an unpredictable event.

The obvious question: ``Given records of my previous measurements 
$|A_{\varphi}\rangle^{\otimes m}$, what are the chances that I, 
the observer, will end up with a record $|A_{k}\rangle^{\otimes l}$ 
of any specific $|s_k\rangle$?" can be now addressed.
The same question can be of course posed and answered in the more general
case, when the confirmation through the re-measurement of the system is not
possible, although then collapse is not as well documented. Here we focus
on non-demolition measurements of hermitean operators to save on notation.

The motivation for considering $|\bar\psi_{\cal SE}\rangle$ or
$|\bar\Psi_{\cal SAE}\rangle$ instead of the pure initial state of the system
in the study of probabilities should be by now obvious: We have already noted
in the course of the discussion of pointer states that the anticipated
and inevitable interaction with ${\cal E}$ will lead from $|\bar\varphi\rangle$
to $|\bar\Psi_{\cal SAE}\rangle$, Eqs. (21)-(23).
As all measurements follow this general pattern, we can -- without any loss of
generality -- anticipate this decoherence - causing entanglement with
${\cal E}$ and start the analysis with the appropriate state.

Observer can in principle carry out measurements of ${\cal S, A}$ or ${\cal E}$
as well as global measurements of ${\cal SAE}$ that will convince him of his
ignorance of the states of individual subsystems and allow him to assign
probabilities to different potential outcomes. The key measurement in this
strategy is the confirmation that the composite systems of interest
are returned to the initial
pure state (say, $|\psi_{\cal SA}\rangle$ or $|\Psi_{\cal SAE}\rangle$) after
the appropriate swaps and counterswaps are carried out. For example,
a sequence of operations:
$$ |A_0\rangle |\bar\psi_{\cal SE}\rangle
{\buildrel {\{\dots \bar\psi_{\cal SE} \dots\}} \over {\longrightarrow}}
|A_{\bar\psi_{\cal SE}}\rangle|\bar\psi_{\cal SE}\rangle \eqno(24a)$$
$$u_{\cal S}(k \rightleftharpoons l) |\bar\psi_{\cal SE}\rangle =
|\bar\eta_{\cal SE} \rangle \eqno(24b)$$
$$ |A_0'\rangle |\bar\eta_{\cal SE}\rangle
{\buildrel {\{\dots \bar\eta_{\cal SE} \dots\}} \over {\longrightarrow}}
|A_{\bar\eta_{\cal SE}}'\rangle|\bar\psi_{\cal SE}\rangle \eqno(24c)$$
$$u_{\cal E}(k \rightleftharpoons l) |\bar\eta_{\cal SE}\rangle =
|\bar \psi_{\cal SE} \rangle \eqno(24d)$$
$$ |A_0''\rangle |\bar\psi_{\cal SE}\rangle
{\buildrel {\{\dots \bar\psi_{\cal SE} \dots\}} \over {\longrightarrow}}
|A_{\bar\psi_{\cal SE}}''\rangle|\bar\psi_{\cal SE}\rangle \eqno(24e)$$
allows the observer to conclude that the even
global state $|\bar\psi_{\cal SE}\rangle$ can be transformed by a swap in
${\cal S}$ into the state $|\bar \eta_{\cal SE}\rangle$, but that it can be
restored to $|\bar\psi_{\cal SE}\rangle$ by the appropriate counterswap in
${\cal E}$. Observers familiar with quantum theory and with the basic
{\bf facts 1-3} concerning the nature of systems that allowed us to establish
Theorems 1 and 2 will be also convinced that probabilities of the outcomes of
the envariantly swappable states of ${\cal S}$ are equal. Sequences of
operations that can be carried out as well as the perfect predictability
of the records (e.g., $\dots |A_{\bar\psi_{\cal SE}}\rangle
|A_{\bar\eta_{\cal SE}}'\rangle |A_{\bar\psi_{\cal SE}}''\rangle\dots)$
that always -- that is, with certainty -- appear at the end of the experiments,
will convince the observer that each of the envariantly swappable outcomes
is equally likely -- that each should be assigned the same probability.

Note that, in principle, observer could treat one of his own record cells as
an external system. This happens when the record is made by the memory of the
apparatus, and the observer is not yet aware of the outcome. Then this record
cell comes to play a role of the extension of the system, i.e., its state
can be swapped along with the state of ${\cal S}$ it has recorded. There is
little difference between this set of measurements and swaps, and the sequence
we have just discussed, Eq. (24). The sequence involving memory cell states
(that are certifiably orthogonal) would allow one to assign probabilities
to the states of ${\cal S}$ that are not orthogonal, as it was noted above.

\subsection{Boolean algebra of records}

We now implement the strategy outlined in Section II, our  `interpretation -
neutral' version of the derivation of Born's rule from envariance, but we
shall use records as `events'.  This will allow us to start at a more fundamental
level than in section II. more. To begin with, we will be able to establish
wnat in section II was the additivity assumption, Eq. (7b). We shall then go on to
derive Boolean logic (which is behind the calculus of probabilities [20]), and
re-derive Born's rule from this `logical' point of view. 

\subsubsection{Additivity of probabilities from envariance}

In the axiomatic formulation of the probability theory due to Kolmogorov 
(see, e.g., Ref. [20]) as well as in the proof of Born's rule due to Gleason [30] 
additivity is an {\it assumption} motivated by the mathematical demand 
-- probability is a {\it measure}. On the other hand, in the standard
approach of Laplace [17] additivity can be established starting from the definition
of probability of a composite event as a ratio of the number of favorable
equiprobable events to the total. The key ingredient that makes this derivation
of additivity possible is equiprobability: We have an independent proof that 
there exists a set of elementary events that are swappable, and, hence, have
the same probability. 

Envariance under swaps is such an independent criterion. Using it we can 
establish {\it objectively} (in contrast to Laplace, who had to rely on the subjective
`state of mind' of the observer) that certain events are equiprobable. We can 
then follow Laplace's strategy and use equiprobability to prove additivity.
This is important, as  additivity of probabilities should not be automatically 
and uncritically adopted in the quantum setting. After all, quantum theory 
is based on the principle of superposition -- the principle of additivity of 
{\it complex amplitides} -- which is {\it prima facie} incompatible with additivity 
of probabilities, as is illustrated by the double slit experiment.

Phases between the record (pointer) states (or, more generally, between any set
of Schmidt states) do not influence outcome of any measurement that can be
carried out on the apparatus (or memory), as Theorem 1 and our discussion
in Section IV demonstrate. This independence of the local state from global phases
invalidates the principle of superposition -- the system of interest (or the
pointer of the apparatus, or the memory of the observer) are `open' 
-- entangled
with the environment. As a consequence, we can, in effect, starting from
envariance, {\it establish} (rather than postulate) Laplacean formula for
the probability of a composite event:

\noindent{\bf Lemma 5}: Probability of a composite (coarse-grained) event
consisting of a subset
$${\bf \kappa} \equiv \{k_1 \vee k_2 \vee \dots \vee k_{n_{\bf \kappa}} \}
\eqno(25)$$
of $n_{\kappa}$ of the total $N$ envariantly swappable mutually exclusive
exhaustive fine-grained events associated with records corresponding to
pointer states of the global state, Eq. (21);
$|\bar\Psi_{\cal SAE}\rangle\ =\ \sum_{k=1}^N e^{i \phi_k}
|s_k\rangle|A_{k}\rangle |\varepsilon_k\rangle\
  =\ \sum_{k=1}^N e^{i \phi_k}
|s_k,A_{k}(s_k)\rangle |\varepsilon_k\rangle$, is given by:
$$ p({\kappa}) = {n_{\kappa} \over N} \ . \eqno(26)$$

To prove additivity of probabilities using envariance we consider the state:
$$ |\Upsilon_{\bar {\cal A}{\cal ASE}}\rangle \ \propto \
\sum_{\kappa} |\bar A_{\kappa}\rangle \sum_{k\in \kappa} |A_k\rangle 
|s_k\rangle
|\varepsilon_k\rangle \eqno(27)$$
representing both the fine-grained and coarse-grained records. We first note
that the form of $|\Upsilon_{\bar {\cal A}{\cal ASE}}\rangle$ justifies
assigning zero probability to $|s_j\rangle$'s that do not appear -- 
i.e., appear
with zero amplitude -- in the initial state of the system. Quite simply, there
is no state of the observer with a record of such zero-amplitude Schmidt states
of the system in $|\Upsilon_{\bar {\cal A}{\cal ASE}}\rangle$, Eq. (27).

To establish Lemma 5 we shall further accept basic implications of envariance:
When there are total $N$ envariantly swappable outcome states, and they exhaust
all of the possible outcomes, each should be assigned probability of $1/N$,
in accord with Eq. (7a). We also note that when a coarse-grained events  are
defined as unions of fine-grained events, Eq. (25), conditional probability of
the coarse grained event is:
$$ p(\kappa | k) = 1 \ \ \ \ k \in \kappa \ , \eqno(28a) $$
$$p(\kappa | k) = 0 \ \ \ \ k \not\in \kappa \ . \eqno(28b)$$
To demonstrate Lemma 5 we need one more property -- the fact that 
when a certain event ${\cal U}$ ($p({\cal U})=1$) can be decomposed 
into two mutually exclusive events, ${\cal U} = \kappa \vee \kappa^{\perp}$, 
their probabilities must add to unity:
$$p({\cal U}) = p( \kappa \vee \kappa^{\perp}) = p(\kappa) +  p(\kappa^{\perp})
= 1\ , \eqno(29)$$
This assumption introduces (in a very limited setting) additivity. It is
equivalent to the statement that ``something will certainly happen''.

{\bf Proof} of Lemma 5 starts with the observation that probability
of any composite event $\kappa$ of the form of Eq. (25) can be obtained
recursively -- by subtracting, one by one, probabilities of all the 
fine-grained
events that belong to $\kappa^{\perp}$, and exploiting the consequences
of the implication, Eq. (28), along with Eq. (29). Thus, as a first step,
we have:

$ p(\{k_1 \vee k_2 \vee \dots \vee k_{n_{\kappa}}\vee \dots \vee k_{N-1}\} +
p(k_N) = 1 \ . $

\noindent Moreover, for all fine-grained events $p(k)={ 1 \over N}$. Hence;

$p( \{k_1 \vee k_2 \vee \dots \vee k_{n_{\kappa}}\vee \dots \vee k_{N-1}\}
= 1-{1 \over N} \ .$

\noindent Furthermore (and this is the next recursive step) 
the conditional probability of the event
$ \{k_1 \vee k_2 \vee \dots \vee k_{n_{\kappa}}\vee \dots \vee k_{N-2}\}$ 
given the event $ \{k_1 \vee k_2 \vee \dots \vee k_{n_{\kappa}}\vee \dots \vee k_{N-1}\}$ is:

$p(\{k_1 \vee k_2 \vee \dots \vee k_{N-2}\}|
\{k_1 \vee k_2 \vee \dots \vee k_{N-1}\})
= 1- { 1 \over {N-1}},$

\noindent and so the unconditional probability must be:

$p(\{k_1 \vee k_2 \vee \dots \vee k_{n_{\kappa}}\vee \dots \vee 
k_{N-2}\}|~ {\cal U}) = (1-{1 \over N})(1-{1 \over {N-1}}).$

\noindent Repeating this procedure untill only the desired
composite event $\kappa$ remains we have:

$ p(\{k_1 \vee k_2 \vee \dots \vee k_{n_{\kappa}}\})=(1 - {1 \over N})\dots
(1- {1 \over {N-(N-n_{\kappa}-1)}}) .$

\noindent After some elementary algebra we finally recover:
$$ p(\{k_1 \vee k_2 \vee \dots \vee k_{n_{\kappa}}\})={n_{\kappa} \over N}\ . $$
Hence, Eq. (26) holds. QED.

\noindent{\bf Corollary}: Probability of mutually compatible exclusive events
$\kappa, \lambda, \mu, \dots$ that can be decomposed into unions of envariantly
swappable elementary events are additive:
$$ p(\kappa \vee \lambda \vee \mu \vee \dots) = p(\kappa) + p(\lambda) + p(\mu)
+ \dots \ \eqno(30)$$

Note that in establishing Lemma 5 we have only considered situations that can
be reduced to certainty or impossibility (that is, cases corresponding to
the absolute value of the scalar product equal to 1 and 0). This is in keeping
with our strategy of deriving probability and, in particular, of arriving at
Born's rule from certainty and symmetries.

\subsubsection{Algebra of records as the algebra of events}

We can take this approach further. To this end, we shall no longer require
coarse-grained events to be mutually exclusive, although we shall continue to
insists that they are defined by the records inscribed in the pointer states.
Algebra of events [20] can be then defined by simply identifying events with
records. Logical product of any two coarse-grained events $\kappa, \lambda$
corresponds to the product of the projection operators that act on the memory
Hilbert space -- on the corresponding records:
$$ \kappa \wedge \lambda \buildrel \rm def \over = P_{\kappa} P_{\lambda}
= P_{\kappa \wedge \lambda} \ . \eqno(31a)$$
Logical sum is represented by a projection onto the union of the Hilbert
subspaces:
$$ \kappa \vee \lambda \buildrel \rm def \over =  P_{\kappa} + P_{\lambda}
- P_{\kappa} P_{\lambda} = P_{\kappa \vee \lambda} \ . \eqno (31b)$$
Last not least, a complement of the event $\kappa$ corresponds to:
$$ \kappa^{\perp} \buildrel \rm def \over = {\bf P_{\cal U}} - P_{\kappa}
= P_{\kappa^{\perp}} \ . \eqno (31c)$$
With this set of definitions it is now fairly straightforward to show:

\noindent {\bf Theorem 4}:
Events corresponding to the records stored in the memory pointer states define
a Boolean algebra.

\noindent{\bf Proof}: To show that the algebra of records is Boolean we need
to show that coarse - grained events satisfy any of the (several equivalent,
see e.g. [42]) sets of axioms that define Boolean algebras:

(a) Commutativity:
$$ P_{\kappa \vee \lambda} = P_{\lambda \vee \kappa}  ;  \ \ \
P_{\kappa \wedge \lambda} = P_{\lambda \wedge \kappa} \ . \eqno(32a,a')$$

(b) Associativity:
$$ P_{(\kappa \vee \lambda)\vee \mu} = P_{\lambda \vee (\kappa \vee 
\mu)}  ;  \ \ \
P_{(\kappa \wedge \lambda)\wedge \mu} = P_{\lambda \wedge (\kappa 
\wedge \mu)} \ . \eqno(32b,b')$$

(c) Absorptivity:
$$ P_{\kappa \vee (\lambda)\wedge \lambda)} = P_{\kappa}  ;  \ \ \
P_{\kappa \vee (\lambda \wedge \kappa)} = P_{\kappa} \ . \eqno(32c,c')$$

(d) Distributivity:
$$ P_{\kappa \vee (\lambda \wedge \mu)} = P_{(\kappa \vee \lambda) 
\wedge (\kappa \vee \mu)}  ;  \ \ \
P_{\kappa \wedge (\lambda \vee \mu)} = P_{(\kappa \wedge \lambda) \vee
(\kappa \wedge \mu)} \ . \eqno(32d,d')$$

(e) Orthocompletness:
$$ P_{\kappa \vee (\lambda \wedge \lambda^{\perp})} = P_{\kappa} ; \ \ \
P_{\kappa \wedge (\lambda \vee \lambda^{\perp})} = P_{\kappa} \ . 
\eqno(32e,e')$$

Proofs of (a)-(e) are straightforward manipulations of projection operators.
We leave them as an exercise to the interested reader. As an example we give
the proof of distributivity:
$P_{\kappa\wedge(\lambda\vee\mu)} =
P_{\kappa}(P_{\lambda} + P_{\mu} - P_{\lambda} P_{\mu}) =
P_{\kappa}P_{\lambda} + P_{\kappa}P_{\mu} - (P_{\kappa})^2P_{\lambda} P_{\mu}
 = P_{\kappa \wedge \lambda} + P_{\kappa \wedge \mu} - P_{\kappa \wedge 
\lambda} P_{\kappa \wedge \mu} =
P_{(\kappa \wedge \lambda) \vee (\kappa \wedge \mu)} \ . $ 
The other distributivity axiom is demonstrated equally easily. QED.

These record
projectors commute because records are associated with the orthonormal pointer
basis of the memory of the observer or of the apparatus: It is impossible to
consult memory cell in any other basis, so the problems with distributivity
pointed out by Birkhoff and von Neumann [43] simply do not arise -- 
when records
are kept in orthonormal pointer states, there is no need for `quantum logic'.

Theorem 4 entitles one to think of the outcomes of measurements --
of the records kept in various pointer states -- in classical terms.
Projectors corresponding to pointer subspaces define overlapping but
compatible volumes inside the memory Hilbert space. Algebra of such composite
events (defined as coarse grained records) is indeed Boolean. The danger of
the loss of additivity (which in quantum systems is intimately tied to
the principle of superposition) has been averted: Distributive law of
classical logic holds.

\subsubsection{Boole and Bohr: interpretational consequences}

In a sense, this means that after a lengthy relative state detour we have
arrived at the resolution of the measurement problem quite compatible with 
the one advocated by Bohr [44]. The set of pointer states of the apparatus 
along with the prescription of how ${\cal A}$ will interact with ${\cal S}$ suffices
to define the ``classical apparatus'' Copenhagen interpretation demands as
a key to the closure of the `quantum phenomenon' brought about by a 
measurement.
Our analysis shows how one can put together such a ``classical apparatus''
from quantum components. Continuous entangling interactions of the memory of
the observer or of the pointer of the apparatus with the environment
are essential. They assure einselection, yielding commuting (and, hence,
Boolean -- see e.g. Ref [45]) sets of events -- sets of future records.

As we have already noted, there is a significant difference between
the transition from quantum to (effectively) classical we have described
above and a measurement on an {\it ab initio} classical system, where one can
always imagine that there is a preexisting answer, waiting to be revealed by
the measurement. In the quantum case the answer is induced by the measurement:
Uncertainty about the outcome arises when the observer who knows the initial
state of the system changes the observable of interest. Once the record has
been made, and the interaction with the environment singled out the preferred
pointer basis, one can act ``as if'', from then on, there was a definite but
unknown outcome -- a probabilistic, effectively classical preexisting event.
Inconsistencies in this neo-Copenhagen strategy based on the existential
interpretation could be exposed only if observer had perfect control
of the environment, and decided to forgo knowledge of the
state of ${\cal S}$ in favor of the global state of ${\cal SE}$.

One of the most intriguing conclusions from the study of the consequences of
envariance for quantum measurements is the incompatibility between the observer
finding out (and, therefore, knowing) the outcome on one hand, and his ability
to expose quantum aspects of the whole on the other: Whenever entanglement is
present, the two are obviously
complementary. Once the state of the observer is described by (and therefore
tied to) a certain outcome, he looses the ability to control global observables
he would need to access to confirm overall coherence of the state vector.
One of the virtues of the existential interpretation is the clarity with which
this complementarity is exhibited.

Specifying measurement scheme along with the coupling to the environment
-- and, hence, with the set of pointer states -- fixes menu of the possible
events. We have already demonstrated (Theorem 4) that such events associated
with future records can be consistently assigned probabilities. The question
that remains concerns the relation between these probabilities and the
pre-measurement state vector. We have arrived at a partial answer:
Whenever the complete set of commuting observables can be fine-grained into
events that are envariantly swappable, the associated fine-grained
probabilities must be equal. We also note that when some potential record
state appears with a coefficient that is identically equal to zero, it must
be assigned zero probability. This is very much a consequence of the 
existential
interpretation: Observer with the corresponding record simply does not exist
in the universal state vector. Last not least, probabilities of mutually
exclusive events are additive.

We conclude that the theory of probability that refers to the {\it records}
inscribed in the pointer states singled out by envariance can be consistently
developed as a classical probability theory. This is in spite of the fact that
quantum states these records refer to cannot be consistently represented by
probabilities. A good example of this situation is provided by the Bell's
inequalities [46]. There the measurement outcomes cannot be predicted by
assuming any probability distribution over the states of the members of
the measured  EPR pair, but records of these outcomes are perfectly classical.
This is because ``an event is not an event untill it becomes a record
of an event'', as one could say paraphrasing John Wheeler's paraphrase [37]
of Niels Bohr. Entangled state of two spins cannot be associated with 
``events''
that correspond to different non-compatible basis states. However, once these
states have been recorded in the preferred basis of the apparatus, event space
can be defined by the records and all the steps that lead to probabilities can
be taken.

\subsection{Born's rule for the probability of a record}

Consider now the case of unequal coefficients, Eq. (8a). After it is recorded
by one of the memory cells of the observer we get:
$$|\psi_{\cal SAE}\rangle = \sum_{k=1}^N \sqrt{m_k/M} |s_k\rangle |A_k\rangle
|\varepsilon_k\rangle \ , \eqno(33)$$
with $M=\sum_{k=1}^N m_k$. Observer can verify that the joint state of
${\cal SAE}$ is indeed $\psi_{\cal SAE}$ by a direct measurement in which one
of his memory cells is treated as an external system. Given this
information (i.e., the form of the joint state, Eq. (33)) he can convince
himself that Born's rule holds -- that it gives correct answers about
the probabilities of various potential records $|A_k\rangle$, and 
corresponding outcome states $|s_k\rangle$ (which need not be 
orthogonal -- see discussion
following Eq. (22)). This can be seen in several different ways [2-4]
of which we choose the following:
Consider another pre-measurement involving a counter ${\cal C}$ that leads to;
$$|\psi_{\cal SEC}\rangle \propto \sum_{k=1}^N \sqrt{m_k}
|s_k\rangle |\varepsilon_k\rangle |C_k\rangle \ \eqno(34)$$
This correlation can be established either by interaction with ${\cal S}$,
or ${\cal E}$ (although this last option may be preferred, as it seems
`safer' -- in absence of any interaction with ${\cal S}$ probabilities of
future records $|A_k\rangle$'s should not change). In any case, the fact
that this can be done by interaction with either ${\cal S}$ or ${\cal E}$
leading to the same $|\psi_{\cal SEC}\rangle$ proves that this
``safety concern'' was not really justified.

As before, we now imagine that:
$$|C_k\rangle = \sum_{j_k=\mu_{k-1}+1}^{\mu_k}
|c_{j_k}\rangle / \sqrt{m_{k}}  \eqno(35) $$
with $\mu_k=\mu_{k-1}+m_k$, and $\mu_0=0$. Note that the observer knows
the initial state of the system $|\varphi\rangle$, Eq. (12a). Hence, we can
safely assume that he also knows coefficients of the state $\psi_{\cal SE}$ or
$\psi_{\cal SAE}$ he will be dealing with. Therefore, finding ${\cal C}$
with the desired dimensionality of the respective subspaces and correlating it
with ${\cal E}$ in the right way is not a ``hit or miss'' proposition -- it
can be always accomplished using the information in the observer's possession.
It is also straightforward in principle to find the environment degrees
of freedom that would decohere the fine-grained states of ${\cal C}$, so
that the complete state would become:
$$|\Psi_{\cal SEC}\rangle \propto \sum_{j_k=1}^M
|s_{k(j_k)}, c_{j_k}\rangle |e_{j_k}\rangle  \eqno(36)$$
where $k(j)=k$ iff $\mu_{k-1} < j \le \mu_k$. This state is obviously envariant
under swaps of $|s_{k(j_k)}, c_{j_k}\rangle$. Hence, by Eq. (7a);
$$ p_{j_k} \equiv p(c_{j_k}) \equiv  p(|s_{k(j_k)}, c_{j_k}\rangle )
= 1 / M \ . $$
Moreover, measurement of the observable with the eigenstates $\{|s_k\rangle\}$
yields:
$$|A_0\rangle |\Psi_{\cal SEC}\rangle \propto \sum_{k=1}^N |A_k\rangle
\sum_{j_k=\mu_{k-1}+1}^{\mu_k} |s_{k(j_k)}, c_{j_k}\rangle |e_{j_k}\rangle
\ . \eqno(37)$$
These disjoint sets correspond to different record states 
$|A_k\rangle$ that are
labled by $k$, each of them containing $m_k$ individual equiprobable events.
Therefore, using Eq. (26), Lemma 5, and its Corollary, we recover Born's rule:
$$p_k \equiv p(s_k) =  \sum_{j_k=\mu_{k-1} + 1}^{\mu_{k}}
p(|s_{k(j_k)}, |c_{j_k}\rangle ) = m_k / M  = |a_k|^2 \ \eqno(38)$$
As before, extending it to the case of when the probabilities are
not rational is straightforward since rational numbers are dense among
the reals.

This appeal to continuity can be made more precise now, providing we
recognize that -- essentially as a consequence of Lemma 5 -- probability of
an event $\lambda$ that includes event $\kappa$ must be at least as large
as the probability of $\kappa$:
$$ \kappa \in \lambda \ \Rightarrow \ p(\kappa) \leq p(\lambda) \ . \eqno(39)$$
This is easily seen as a consequence of Eqs. (29) and (30). It is then
straightforward to set up a limiting procedure that bounds an irrational
probability from above and from below with sequences of states with rational
probabilities such that
$m_{\nu}^-/M_{\nu} \leq  p(\kappa) \leq  m_{\nu}^+/M_{\nu} \leq $. As $M_{\nu}$
approaches infinity, $p(\kappa)=|\alpha_{\kappa}|^2$ obtains in the limit.

We have waited until now with detailing this continuity argument because 
it can be rigorously put forward only after additivity of probabilities has been
established. And, as was known since the inception of modern quantum theory,
superposition principle is in conflict with additivity of probabilities:
For example, Eq. (39) would {\it not} hold if `events' were not associated with
the records, as Eq. (30) does not hold for arbitrary states.  Furthermore,
Eq. (30) and the distributivity axiom are violated in double slit experiment
if `particle passing through the left (right) slit' are identified as events.
They regain validity when `the (pointer basis) record of a particle passing
through the left (right) slit' are regarded as events. We note again that this is
essentially Bohr's mantra, as reported by Wheeler [37]! The only difference
is that the record need not be made by an {\it ab initio} a classical apparatus
-- an effectively classical apparatus with a set of memory states fixed as 
a consequence of incessant monitoring by ${\cal E}$ suffices to do the job.
It takes a bit more effort to extend Born's rule to the case of continuous spectra.
We show hoe to do that in Appendix A.

In this section we have established that events associated with anticipated
future records correspond to a Boolean structure. This allowed us to assign
probabilities to potential outcomes. Any compatible set of outcomes
-- any set of measurements that can be associated with orthonormal memory
states of the observer, apparatus, etc. -- can be analysed in this fashion.
Observer makes a choice of what he will measure, but the inevitable
entangling interaction with the environment will select a certain preferred set
of pointer states. So, ultimately, observer uses probabilities to anticipate
his future state.

Once we have established that pointer states can be assigned probabilities, we
have asked about their connection with the coefficients of the pre-measurement
state of the system. Here the answer was based on the same idea of envariance
invoked in Section II. In effect, coarse-grained outcome states are compatible
and can be always fine-grained by using suitable ancilla so that 
Hilbert - space volumes corresponding to various possible compatible 
fine-grained outcomes
contain same ``concentrations'' of probability. This was done by ``dilution''
of the original state with the help of the counterweight - ancilla ${\cal C}$. Envariance
can be then used to confirm that all of the fine-grained cells must be assigned
the same measure -- and hence, the same probability. The probability of
coarse-grained events was derived by counting the number of fine-grained cells.
It is given (as we have established in Lemma 5) by the fraction of the total
number of such envariant (and, hence, equivalent) cells. This is very much in
the spirit of Laplacean definition of probability -- ``ratio of the number of
favorable events to the total number of events''. The advantage of the quantum
discussion rests in its ability to rigorously show when such elementary events
are envariantly swappable, and, hence equiprobable. This transforms a
subjective definition based on the state of mind of the observer into objective, 
experimentally testable, statement about symmetries of entangled states.

\section{Born's rule, relative frequencies, and envariance}

Envariance - based derivation of Born's rule introduces probability as a tool
observer adopts to predict the future -- or, more precisely, to predict his future state
given that he decides to measure some specific observable. Outcomes of planned
measurements are uncertain because of quantum indeterminacy. Even when observer
knows all that can be known about the system -- even when ${\cal S}$ is in
a pure state -- ignorance appears whenever the to - be - measured observable
does not have the state prepared by the preceding measurement among
its eigenstates.

Observers aim is to assess likelihood of a particular future record in
comparison with the alternatives. The measure that emerges is based on
the equiprobability of certain mutually exclusive events (orthogonal states) under
swaps. They are provably equally probable because the global ${\cal SE}$ 
state can be restored by counterswaps in the environment. Environment 
can be invoked already for pure state of ${\cal S}$, Eq. (12), when 
contemplating possible outcomes of the future measurements: it will 
inevitably and predictably entangle with the records causing decoherence 
and unpredictability.

The fundamentally predictive role of probabilities reflected in our
derivation of Born's rule is often contrasted with the ``relative frequency
interpretation'' [18,19]. Imagine an observer who -- instead of counting
numbers of mutually exclusive envariantly swappable states -- performs
the same experiment over and over, and infers his chances of getting a certain
outcome in the next round from the past records, by assuming that he is in
effect dealing with an infinite ensemble deduced from his finite data.

Supporters of the Bayesian / Laplacean `epistemic'  and subjective view 
of probabilities and of the opposing `relative frequency' approach have
been often at odds, taking their own views to dogmatic extremes while
pointing out flaws of the opposition [19]. The central difficulty pointed out
by the frequentists in their criticism of the Bayesian approach -- that
ignorance gives one no right to make any inferences, and hence, no
right to assign {\it any} probabilities to the possible outcomes -- is difficult
to ignore. Therefore, Bayesian appeals to symmetry through the principle 
of indifference are inappropriate when they involve `state of mind' of the observer. 

However, while such criticisms are very relevant in the classical setting, they
simply do not apply here: In quantum physics ignorance of a future outcome 
can be demonstrated and quantified by employing objective symmetries of
the preexisting state (that can be perfectly known to the observer). This was 
our strategy. Thus, in questions of fundamental significance it would seem 
appropriate to deduce probability by identifying the relevant quantum symmetry --
envariance under swaps -- and by counting the fundamentally swappable (and,
hence, equiprobable) outcomes. 

This strategy may not be always applicable. For instance, in some
situations where such fundamentally equiprobable events are difficult
to identify. So, reliance on actuarian tables in the insurance business is 
difficult to question. However, in quantum measurements we are dealing 
with probabilities of a single event at a very fundamental level. Frequencies 
should be secondary. Relative frequency approach was rightly criticized for 
requiring infinite ensembles. Indeed, this task of extrapolating -- deducing infinite 
ensemble required for future predictions from the relative frequencies of the past
outcomes -- involves a subjective element studied e.g. by de Finetti [47].

Nevertheless, even when one can deduce probabilities {\it a priori} using
envariance, they better be consistent with the relative frequencies estimated
by the observer {\it a posteriori} in sufficiently large samples. Such
a ``consistency check'' is one of the motivations for this section.
More importantly, our discussion will explore and clarify the relation
of the {\it a priori} probabilities with the relative frequencies. This has
significance for the understanding of the implications of envariance in the
existential interpretation of quantum physics. We shall conclude that when 
probabilities can be deduced directly from the pure state, the two approaches 
are in agreement, but that the {\it a priori} probabilities obtained from 
envariance - based arguments are more fundamental.

Given the central idea of our approach -- that the symptoms of classicality and
the effective collapse are induced by the interaction with the environment --
we shall conduct our discussion in the relative state setting. Thus, we imagine
observer presented with a large ensemble of systems, each prepared in the same
state $|\varphi\rangle$, Eq. (12). Observer employs a counter ${\cal C}$ to
(pre-)measure each of the systems in the same basis $|s_k\rangle$:
$$|C_0\rangle^{\otimes {\cal N}} |\varphi\rangle^{\otimes {\cal N}}  \ \
{{\buildrel {\{|s_k\rangle\}} \over
{\underbrace {\Longrightarrow \dots \Longrightarrow}_{\cal N}}}} \dots $$
$$ \dots \ \prod_{l=1}^{\cal N} \Bigl(\sum_{k=1}^N \langle s_k|\varphi\rangle
|C_k\rangle |s_k\rangle \Bigr)_l \ = |\Phi_{\cal SC}^{\cal N}\rangle\ .
\eqno(40)$$
Note that above ${\cal N}$ measurements are carried out `in parallel', one on
each of ${\cal N}$ systems. We have reflected this difference in notation by
using ``$\Longrightarrow$'' instead of ``$\longrightarrow$'' of, say, 
Eq. (11b).
To simplify the notation we limit our considerations to the case when
there are just two possible outcomes, so that;
$$|s_1\rangle = |0\rangle \ , \ \  |s_2\rangle = |1\rangle \ , $$
with $\langle 0 | 1 \rangle = 0$. Moreover, as before, we imagine that for some
integer $m$ and $M$ ($0 < m < M$) we have
$ |\alpha|^2 = |\langle 0 | \varphi \rangle|^2 = {m / M}, \
| \alpha|^2 + |\beta|^2 = 1$, so that a countable way exists to further
``resolve'' the state above into superpositions that have the same absolute
values of the coefficients. In this manner, starting from
$$ |\varphi\rangle = \alpha |0\rangle + \beta |1\rangle  \eqno(12c)$$
we arrive at
$$|\Phi_{\cal SCE}^{\cal N}\rangle \ \propto \
\prod_{l=1}^{\cal N} \bigr(\sum_{j=1}^m |0\rangle |c_j\rangle |e_j\rangle  +
\sum_{j=m+1}^M |1\rangle |c_j\rangle |e_j\rangle\bigl)_l \ .
\eqno(41)$$
The steps that lead to this state in the composite Hilbert space $({\cal H_S}
\otimes {\cal H_C} \otimes {\cal H_E})^{\otimes {\cal N}}$ can be reproduced by
the reader following the strategy of Section II.
We assume (in analogy with Eqs. (9) \& (35)) that:
$$|C_0\rangle = \sum_{j=1}^m |c_j\rangle /\sqrt m \ ;
\ \ |C_1\rangle = \sum_{j=m+1}^M |c_j\rangle /\sqrt {M - m} \ .$$
In effect, we assume that the Hilbert space of each counter cell has
a sufficient dimensionality to allow for the increased resolution needed to
``even the odds'' between the mutually exclusive fine-grained $|c_j\rangle$.

We can now carry out the product in Eq. (41). The resulting
sum has $M^{\cal N}$ distinct terms:
$$|\Phi_{\cal SCE}^{\cal N}\rangle \ = \ M^{-{\cal N}} \sum_{\{h^{\cal C}\}}
|h^{\cal SCE}\rangle \ . \eqno(42a)$$
Individual recorded histories have the form of ordered products:
$$ |h^{\cal SCE}\rangle = \dots
(|0\rangle_{k_1} |c_{j\leq m}\rangle_{k_1} |e_j\rangle_{k_1}) \dots
(|1\rangle_{k_2} |c_{j> m}\rangle_{k_2} |e_j\rangle_{k_2})
\dots \eqno(43a)$$
That is, they reflect individual sequences of fine - grained records made by
the observer. Each $|h^{\cal SCE}\rangle$
is completely determined by the history of fine - grained counts:
$$|h^{\cal C}\rangle\ = \ \dots  |c_{j}\rangle_{k_1} \dots
|c_{j}\rangle_{k_2}  \dots \ \  \eqno(43b)$$
Thus, $h^{\cal SC}$ is obviously implied by $h^{\cal C}$. Moreover, by
assumption, states of the ${\cal N}$ distinct environments are individually
Schmidt, Eq. (41), that is orthonormal and in one - to - one 
correspondence with
the states of ${\cal C}$. So, in the end, given the initial state of 
${\cal E}$,
each $h^{\cal SCE}$ is completely determined by
$h^{\cal C}$: The history of the fine-grained counts implies the whole history
of measurements -- sequences of the detected states of the system as well as
the `history of decoherence':
$$|h^{\cal SCE}\rangle = 
\prod_{k=1}^{\cal N} (|s_k(c_j)\rangle |c_j\rangle |e_j\rangle)_k
 = |h^{\cal S}(h^{\cal C}) \rangle \otimes |h^{\cal C}\rangle \otimes
|h^{\cal E}(h^{\cal C})\rangle \ . \eqno(44)$$
This structure of the set of complete histories and the fact that
$h^{\cal C} \Rightarrow h^{\cal SC}$ as well as
$h^{\cal C} \Rightarrow h^{\cal SCE}$ are important for two reasons. We shall
use it to prove that the superensemble $ |\Phi^{\cal N}_{\cal SCE}\rangle\ $
is suitably envariant (so that we can attribute the same probability to
the distinct histories $h^{\cal SC}$). Moreover, we shall have to count
numbers of equiprobable histories that yield the same numbers of detections
(of, say, ``1'') in ${\cal S}$ to compute relative frequencies.

Let us start with the proof of envariance. We need to demonstrate that any two
fine-grained histories $h^{\cal SC}$ that appear in the above superensemble,
Eq. (44), can be envariantly swapped. A general form of the history 
representing
correlated states of ${\cal S}$ and ${\cal C}$ only is:
$$ |h^{\cal SC}_{j_1j_2...j_{\cal N}}\rangle \ =
\ |s_1\rangle |c_{j_1}\rangle ~|s_2\rangle |c_{j_2}\rangle
  \dots |s_k\rangle |c_{j_k}\rangle \dots
|s_{\cal N}\rangle |c_{j_{\cal N}}\rangle \ \eqno(45)$$
The ${\cal SC}$ swap operator that exchanges any two histories is given by:
$$ u_{\cal SC}(h^{\cal SC}_{j_1j_2...j_{\cal N}}\rightleftharpoons
h^{\cal SC}_{j_1'j_2'...j_{\cal N}'})  \ = \
|h^{\cal SC}_{j_1j_2...j_{\cal N}}\rangle \langle h^{\cal 
SC}_{j_1'j_2'...j_{\cal N}'}|
+ h.c. \eqno(46a)$$
or, more succinctly;
$$ u_{\cal SC}(h^{\cal SC} \rightleftharpoons h'^{\cal SC})
\ = \ |h^{\cal SC}\rangle \langle h'^{\cal SC}| + h.c. \ \eqno(46b)$$
When both histories appear in the state, Eq. (41), this swap can be obviously
undone by a counterswap in the environment:
$$ u_{\cal E}(h^{\cal E}_{j_1j_2...j_{\cal N}}\rightleftharpoons
h^{\cal E}_{j_1'j_2'...j_{\cal N}'})
= |e_{j_1}\rangle...|e_{j_{\cal N}}\rangle
\langle e_{j_{\cal N}'}| ... \langle e_{{j_1}'} |
+ h.c. \eqno(46c)$$

We have now established that each fine-grained history in our many - worlds
version of an ensemble has the same probability:
$$ p(h) = M^{-{\cal N}} \ . \eqno(47)$$
To facilitate the calculation of the relative frequencies we sort superensemble
of Eq. (42a) which has $M^{\cal N}$ terms into $2^{\cal N}$ terms that differ
only by the pattern of detections of the state ``$|1\rangle$'' (i.e., we group
different fine-resolution terms together). Next, we group these $2^{\cal N}$
terms into ${\cal N}+1$ terms that differ only in the total number $n$ of 1's
(i.e., we ignore the ``order of appearance'' of 0's and 1's).

This operation involves summing up numbers of detections of 1's. It could even
be implemented in the memory of the apparatus or of the observer by a
{\it register} $\Re$. Register performs a unitary transformation --
a `subroutine' that operates on states of the counter ${\cal C}$ that has
the record of the complete fine-grained history. Register sums up the number of
detections of 1, and writes down the result ``$n$'' in the suitable register
cell:
$$\Re |0\rangle_{\Re} |h^{\cal C}\rangle = |n\rangle_{\Re} |h^{\cal C}\rangle
\ . $$
Register cell could work equally well by summing up the number of 1's
directly in the states of the system ${\cal S}$, Eq. (43a), but it seems
more appropriate to let the observer use ${\cal C}$ for this purpose.
Register cell has at least ${\cal N}$ possible states.

The result can be written in the abbreviated from as:
$$|\Phi_{{\cal SC} \Re {\cal E}}^{\cal N}\rangle \  \propto
\ \sum_{h^{\cal C}} |n\rangle_{\Re}|h^{\cal SC}\rangle |h^{\cal E}\rangle \ =
\eqno(42b)$$
$$ \sum_{n=0}^{\cal N}
{{\cal N} \choose n} |n\rangle_{\Re} |0\rangle^{{\cal N}-n}
|1\rangle^n (\sum_{j=1}^m |c_j\rangle |e_j\rangle)^{{\cal N}-n}
(\sum_{j=m+1}^M |c_j\rangle |e_j\rangle)^n . \eqno(42c)$$
The first line above has all the $M^{\cal N}$ histories -- it represents the
whole superensemble before any sorting was implemented.  Eq. (42c) groups
histories with the same numbers of 1's together. There, we have compensated for
not distinguishing between ${{\cal N} \choose n}$ distinct
sequences of 0's and 1's with the coefficient.

The probability that the observer will detect $n$ of 1's in ${\cal N}$
measurements is proportional to the number of envariantly swappable
fine-grained histories with $n$ detections of 1:
$$p_{\cal N}(n) = {{\cal N}\choose n}{{m^{{\cal N}-n}(M-m)^n}\over M^{\cal N}}
= {{\cal N} \choose n} |\alpha|^{2({\cal N}-n)} |\beta|^n \ . \eqno(48)$$
This is of course the familiar binomial distribution. As before, we can address
the case when probabilities are not commensurate by noting that rational
numbers are dense among reals. Generalisation to the case when there are
more than two outcomes is straightforward.

The above discussion was valid for any ${\cal N}$. We now note that, for
a large ${\cal N}$ binomial distribution of Eq. (48) can be approximated
by a Gaussian:
$$ p_{\cal N}(n) \ = \ {{\cal N} \choose n}
|\alpha|^{2({\cal N}-n)} |\beta|^n $$
$$ \approx \
{1 \over {\sqrt { 2 \pi {\cal N}} |\alpha \beta|}} \exp \{ -
\biggr({{n- {\cal N} |\beta|^2} \over {\sqrt {\cal N} |\alpha \beta 
|}} \biggl)^2\} \ . \eqno(49)$$
Hence, in the limit of large ${\cal N}$ relative frequency of 1's is sharply
peaked around the value predicted by the Born's rule:
$$ r_{Born} \ = \ { n_{Born} \over {\cal N}} \ = \ |\beta|^2 \ . \eqno(50)$$
The peak has a finite width, so that the expected deviation $\delta n$ is
of the order:
$$ \delta n \ = \ \sqrt{\cal N} |\alpha \beta| \ . \eqno(51)$$
Generalization to the case when there are more than two outcomes is
straightforward: All of the steps leading to Eq. (48) can be repeated, and
the large - ${\cal N}$ limit can be the taken using Tchebychev's theorem [20].

While the relative frequency $r=n/{\cal N}$ is sharply peaked around
$r_{Born}$, its `correct' value, the probability that the number of 1's
concides {\it exactly} with $n_{Born}$ predicted by Eq. (50) tends to
{\it decrease} with increasing ${\cal N}$. This is so even when $|\beta|^2$ is
a rational number with a denominator that is a multiple of ${\cal N}$. More
significantly, the count of envariantly swappable histories that yield
$n \in [n_{Born}-\Delta n, n_{Born}+\Delta n]$ decreases with increasing
half-width of the distribution, Eq. (51), as 
$\Delta n/\sqrt{\cal N}|\alpha\beta|$.

On the other hand, half-width of the distribution increases only with
$\sqrt{\cal N}$. Therefore, for a sufficiently large ${\cal N}$ almost all
histories will yield relative frequency within any fixed size relative
frequency interval $r_{Born} - \Delta r <r< r_{Born} + \Delta r$. Moreover,
$$ \lim_{{\cal N} \rightarrow \infty} p_{\cal N}(n / {\cal N}) \ = \
\delta (r - r_{BORN}) \ . \eqno(52)$$
Analogous conclusions are known in the standard probability theory [20].
We recount them here for the quantum superensemble, Eq. (42), as similar
questions have led to confusion in the past in the quantum setting of Many
Worlds relative frequency derivations (see e.g. the critical comments of
Squires [28] pointing out inconsistencies of the frequency operator - based
approaches [22,23] to the derivation of Born's rule).

It is tempting to compare superensemble of Eqs. (42) and (44) with 
the `collective' employed (especially by von Mises [18]) to define
probabilities using relative frequencies. Collective is an ordered infinite 
ensemble of events when; (i) it allows for the existence of limiting relative 
frequencies that are (ii) independent under a selection (using 
so-called `place selection function')
of any infinite subset of the members of the collective. Place selection
functions are meant to represent betting strategies: Player (or obsrever)
can select (or reject) the next member of an ensemble using any algorithm
that does not refer to the outcomes of future measurements -- to the state
of the next member of the ensemble -- but only to its `adress', its `place'
in the ensemble. There was initially some controversy related to
this randomness postulate: Obviously, it is possible to influence limits in
infinite series by selecting subsets of terms. This problem has been however
settled (more or less along the lines anticipated by von Mises) by
the work of Solomonoff, Kolmogorov, and Chiatin: Algorithmic randomness they
have introduced (see [48] for overview and references) provides a deep and
rigorous definition of what is random. However, the very idea of using
{\it infinite} ensembles to define probabilities forces one to extrapolate
{\it finite} data sets -- outcomes of measurements. Any such extrapolation is
a subjective guess.

Chance and determinism combine in an unanticipated manner in quantum theory:
Everettian superensemble of Eq. (42) evolves unitarily, and, hence
deterministically.  Nevertheless, it satisfies a natural generalization of
the von Mises' randomness postulate without any restrictions on place
selection functions. This is easy to see, as each measurement yields every
outcome with the same coefficients, so choosing any subset of measurements
-- any subset of $k$'s in the product representation of the superensemble
-- will have no effect on the limiting relative frequencies.

The randomness postulate of von Mises is mirrored by the assumption of
`exchangeability' introduced in a fundamentally very different (`subjectivist')
approach based on decision theory (see e.g., de Finetti [47]). There the goal
is to deduce probability of the next outcome from a finite sequence of
outcomes of preceding measurements. Exchangeability is meant to assure that
`rules of the game' do not change as the consecutive measurements are carried
out. It captures the (von Mises') idea that each new measurement is `drawn
at random from the same collective' but -- in keeping with the Laplacean
spirit -- avoids reference to a pre-existing infinite ensembles.
Superensemble of Eqs. (40)-(42) is obviously exchangeable:  its is 
a superposition of all possible ensembles, with all possible relative
frequencies. Hence, the order of any two measurements can be permuted 
with no effect on the partial relative frequencies, etc., providing that 
the preparation of the initial state vector and the measured observable 
remain unchanged. 

`Maverick branches' with relative frequencies that are inconsistent with Born's
rule (e.g., a branch with 1's only) plagued Many Worlds relative frequency
approach [21-28]. Maverick branches are `alive and well' in the superensemble,
but have negligible probabilities (deduced now directly from envariance) and,
therefore, for ${\cal N} \gg 1$, are of little consequence.  We have
already discussed the case of small departures above. As the number
${\cal N}$ of measurements increases, probability of detecting a frequency
that is inconsistent with the predictions of Born's rule becomes negligible
in accord with Eqs. (42), (48) and (49). Thus, it is {\it conceivable but very improbable} 
that observer will record in his experiments relative frequency of 1's that is 
far from $r_{Born}$.

\section{Discussion}

Past approaches to the derivation of Born's rule have often had, 
usually as an explicitly stated  goal,  `recovery'  of the classical 
definition probabilities. Limitations inherent in such
a formulation of the problem were in part responsible for their
limited success. `Recovery' of someting as ill-defined and controversial
as any of the classical definitions of probability was bound to be
plagued with difficulties. Moreover, `recovery' of classical probabilities 
suggests that the process should start with getting rid of the most 
quantum aspects of quantum theory to make it closer to classical. 

The only known way to recognize effective classicality in a wholly 
quantum Universe is based on decoherence. But decoherence is 
`off limits' as it employs tools dependent on Born's rule. On the 
other hand, when classicality was `imposed by force' by Gleason [30] 
or in Refs. [31,32], this seemed to work to a degree, although 
interpretational issues were left largely unaddressed and doubts 
have rightly persisted [26-29,33].

\subsection{Envariance and decoherence}

We have taken a very different approach. Derivation of Born's rule described
here is extravagantly quantum: Envariance relies on entanglement, perhaps
the most quantum manifestation of quantum physics, still regarded by some as
a `paradox'. Instead of first taming quantum theory to make it look classical, 
we have used purely quantum symmetries of entangled states as a key 
to unlocking the meaning of probability in a quantum Universe. So, to
arrive at Born's rule we have put aside (at least temporarily) tools -- and
hence, results -- of decoherence, as relying on them threatened circularity. 
This meant that even such basic symptoms of classicality and key ingredients 
of contemporary quantum measurement theory as the einselection of preferred
pointer states had to be motivated and re-derived anew.

Putting aside tools and results of decoherence did not force us to forget about
the role of the environment. To understand emergence of the classical one must
regard quantum mechanics as theory of correlations between systems. This 
stresses the relational aspects of quantum states emphasized
by Rovelli [49]. Moreover, to understand the origins of ignorance, to motivate 
introduction of probabilities, and to appreciate their role in making predictions
one needs to analyse systems that interact with their environments, and focus
on correlations that survive in such open settings.

Once these (in retrospect, natural) steps are taken, it is possible to see how 
observer can be ignorant about the outcome of the measurement he is about
to perform on the perfectly known system -- on ${\cal S}$ that is in a pure state:
Ignorance of the future outcome arises as a consequence of quantum 
indeterminacy and the interaction (of the system, the pointer of apparatus, or of
the observers memory) with the environment. Decoherence inevitably follows
pre-measurements. Entanglement between the system and the memory takes memory
out of the `ready to measure' pointer state into a superposition of (outcome)
pointer states. This in turn means that the ${\cal SA}$ entanglement spreads
into the correlations with the environment in such a way that the observer may
as well assume from the outset that ${\cal S}$ (and / or ${\cal A}$) was
entangled with ${\cal E}$.

Eigenstates of the to - be - measured hermitean observable turn -- as a result
of measurement and decoherence -- into Schmidt states of ${\cal S}$ 
in the resulting decomposition of the entangled state of ${\cal SE}$. 
This is also the case when ${\cal E}$ is in a mixed state to begin with 
-- such states can be purified, and conclusions about the Schmidt states 
of ${\cal S}$ follow. The rest of our argument goes through unimpeded. 
Thus, observer may as well recognise the inevitable and focus on 
the remaining information about the system that can be deduced from 
the resulting entangled state. Envariance can be then readily demonstrated
and -- given additional assumptions we shall not recapitulate here -- used to
arrive at Born's rule. And when measurements are not ideal, (i.e., do not
preserve the state of the system, do not correlate record states with
orthonormal states of ${\cal S}$, etc.) future record states of ${\cal A}$
(which can be safely assumed to be orthogonal) can be used to motivate
envariance -- based approach.

One key difference between the classical definition of probability 
and the quantum view of the envariant origin of ignorance and our
derivation of Born's rule for probabilities is the reliability of the
prior information about the state of the system available to the observer
in the quantum case: That is, observer can use his information about
the initial state of the system $|\varphi\rangle$ to assess chances
of outcomes of future measurements on ${\cal S}$ he may contemplate.
In classical discussions the nature and the implications of ignorance were
the most contentious issues [17-20]. Infering prior probability
distribution `from ignorance alone' is impossible, as was rightly noted by
frequentists. Attempts to invoke symmetry [50] were ultimately unconvincing
[19], as they had to refer to subjective knowledge of the observer, and not
to the underlying physical state.

By contrast, in quantum physics observer can reliably deduce the extent of his
ignorance about the future outcome from the perfect information he has about
the present state of the system, and verify his symmetry arguments by
performing appropriate swaps and confirmatory measurements on the state
of e.g., the composite ${\cal SE}$. This approach preserves some of the spirit
of Laplace, but now the analogue of `indifference' is no longer subjective:
It is grounded in measurable -- and testable if not yet deliberately tested
-- quantum symmetry of entangled states.

Once inevitability of correlations with ${\cal E}$ is recognised, their key
consequence -- environment - induced superselection -- can be recovered
{\it without} usual tools of decoherence: Trace and reduced density matrices
can be put aside in favor of more fundamental approach based on correlations
and environment - assisted invariance. Envariance shows that phases of states
appearing in the Schmidt decomposition are of no consequence for results of any
measurement on the subsystem ${\cal S}$ (or, for that matter, ${\cal E}$) of
the whole ${\cal SE}$. So, superpositions of Schmidt states of ${\cal S}$ cannot
exist, as relative Schmidt phases have no bearing on the state of ${\cal S}$.

\subsection{Envariance behind the ignorance}

Envariance pinpoints the source of ignorance: It is ultimately traced to
the global (quantum) symmetry of entangled states which in turn implies
non-local character of quantum phases of the coefficients in the Schmidt
decomposition. This immediately leads to the envariance of swaps -- they are
generated by changing phases in the eigenvalues of unitaries diagonal 
in a basis complementary (e.g., Hadamard) to the eigenstates of the 
to - be - measured hermitean observable. Swaps are envariant when 
the state is even and the corresponding Schmidt
coefficients have the same absolute values: Non-locality of phases implies
ignorace about states of subsystems, and, hence, of the outcome. Once
envariance of phases is accepted, envariance under swaps follows and implies
equiprobability. A simple counting argument leads then very naturally to Born's
rule in case when coefficients have unequal absolute values.

Note that even an observer presented with an unlimited supply of copies of the
``${\cal S}$ half'' of identically prepared ${\cal SE}$ pairs will eventually
conclude -- having carried out all the conceivable measurements -- that the state
of ${\cal S}$ is mixed, and can be represented by the usual reduced density
matrix. This is a significant difference from other approaches, which do not
recognize the role of the environment. It further underscores the objective
nature of probabilities implying ignorance about the future outcomes. Similar
level of objectivity  is impossible to attain in approaches that do not
recognize role of the environment in the quantum - classical transition
precipitated by quantum measurements. In particular, in absence of entanglement
with some ${\cal E}$ purity of the underlying state of ${\cal S}$ would make
it impossible to apply arguments based on permuting outcomes
to pure states of the system: Such permutations generally change the state
of the system, and can be detected. As the local state is altered, there is no
compelling reason to assume that probabilities would remain unchanged.

An example of an approach threatened by this difficulty is Wallace's
elaboration [36] of an idea -- due to Deutsch [24] -- to apply decision
theory to pure states in order to arrive at Born's rule. In the original
paper [24] this difficulty was not obvious as the presentation left open the
possibility of a ``Copenhagen" point of view with explicit collapse, where
phases do not matter. Indeed, early criticism by Barnum et al. [51] was based
on inconsistencies implied by the `Copenhagen reading' of Ref. [24]. Wallace
has pointed out in a series of more recent papers that that criticism does not
apply to the ``Everettian reading'' of [24]. The approach of Ref. [36]
is however open to two separate charges. Reliance on the (classical) decision
theory makes the arguments of [24] and [36] very much dependent on decoherence
as Wallace often emphasizes. But as we have noted repeatedly, decoherence
cannot be practiced without an independent prior derivation of Born's rule.
Thus, Wallace's arguments (as well as similar `operational approach' of
Saunders [52]) appear to be circular. Even more important is the second
problem; permuting potential outcomes (e.g., using swaps to change
$|\alpha\rangle \propto |1\rangle +|2\rangle-|3\rangle+|4\rangle$ into
$|\beta\rangle \propto |1\rangle +|3\rangle-|2\rangle+|4\rangle$) {\it changes}
the state of an isolated system. And a different state of ${\cal S}$ could
imply different probabilities. So the key step -- irrelevance of the phases for
probabilities of the outcomes (which we have demonstrated in Theorem 1 by
showing that state of ${\cal S}$ is unaffected by envariant transformations)
-- cannot be established without either relying on ${\cal E}$ or some very
strong assumptions that would have to, in effect, invalidate the principle of
superposition.

Early assesments of envariance - based derivation of Born's rule [13,16,53]
are incisive but also generally positive. They have focused on 
the equal coefficients part of the proof, covered here by Theorems 1 and 2. 
This is understandable, as the proof of equiprobability from envariance
is  the key to the rest of the derivation. The consensus so far seems to be
that, apart from the need to clarify some of the steps (the task we undertook 
here) the envariant derivation [2-4] of Born's rule stands.
Indeed, as argued by Barnum, assumptions of the original derivation could be
relaxed by exploiting consequences of envariance more completely [13].
Moreover, there are several interesting and non-trivial variants of the original
proof [3,13,16], which suggests that envariance-based derivation is robust,
and that the vein of the physical intuition it has tapped is far from exhausted.

This paper was also written as a critical if (understandably) friendly review 
and extension of Refs. [2-4], although with a different focus: Any attempt
at the derivation of Born's rule must be completely independent of any and 
all of its consequences we have come to take for granted. Therefore, 
my focus here was to look for circularity, and to make certain that there is none
in the derivation. In this spirit, I have fleshed - out the `decoherence - free' 
definition of pointer states that was briefly discussed in Refs. [2] and [3]. 

Pointer states are the key ingredient of the quantum measurement theory. 
They define the alternatives in the `no collapse' approach to measurement
-- they are the potential events observer is going to place bets on using 
probabilities. Their existence was demonstrated using envariance - inspired 
argument -- by exploring stability of correlations (e.g, apparatus-system) in 
the `open' setting -- in presence of envariance.

We have noted the dilemma of the observer who can either settle
for the perfect knowledge of the whole -- of the global state which is useless
for most purposes but could in principle allow for reversibility -- or opt to
find out the outcome of his measurement, which will irrevocably `tie him down'
to the branch labelled by the outcome. These existential consequences of the
information acquisition would preclude him from reversing the measurement.

Noncommutativity of the relevant global and local observables is then ultimately
responsible for the observer's inability ``to be a Maxwell's demon'' -- to find
out the outcome while retaining the option of reversing the evolution.
Thus, envariance sheds a new light on the origins of the second law in the
context of measurements, complementing ideas discussed to date (see e.g.
[54-56]) and may be even relevant to some old questions concerning
verifiability and consistency of quantum theory [57]. 

In particular, complementarity of global and local information emphasizes 
the difference between attempting reversal
in classical and quantum settings: In classical physics the state of the whole
is a Cartesian product of the states of parts. Hence, in classical physics
perfect knowledge of the state of the whole implies perfect knowledge of all
the parts. In quantum physics only a subset of states of measure zero in the
composite Hilbert space -- pure product states -- allow for that. In all other
cases (which are entangled) knolwedge of the local state (i.e., of ${\cal S}$
precludes knowledge of the global state (i.e., of ${\cal SA}$ or ${\cal SE}$).
But it is the global state that is needed to implement a reversal. We shall
pursue this insight into the origins of irreversibility elsewhere.

\subsection{Remaining questions and future research}

Our study of the quantum origins of probability has not addressed all of 
the questions that can and should be raised. It is therefore appropriate 
to point out some of the issues that will benefit from further study. We start 
by noting the paramount role of the division of the Universe into systems.
Systems are the subject of axioms (o) and (ii), as well as {\bf facts 1-3} 
of Section II. I have signalled this question before [3,8], but as yet there 
has been really no discernible progress towards the {\it fundamental} answer
of what defines a system.

Another interesting issue is the distinction between `proper' and `improper'
mixtures (see e.g. [58]) and the extent to which this may be relevant to the
definition of probabilities. An example of a proper mixture is an ensemble of
pure orthonormal states (eigenstates of its density matrix) mixed in the right
proportions. All that decoherence can offer are, of course `improper mixtures'.
It has been often argued that improper mixtures cannot be interpreted through
an appeal to ignorance [53,58,59]. I believe there is no point belaboring this
issue here: As we have noted before, observer can be ignorant of his future
state, of the outcome of the measurement he has decided to carry out. 

Whether this sort of ignorance is what used to be meant by `ignorance' in 
the past discussion of the origin of probabilities may be of some historical 
interest, but ignorance of the outcome of the future measurement is clearly 
a legitimate use of the concept, and, as we have seen above, quite fruitful.
In a sense we have touched on an issue related to the distinction
between proper and improper mixtures when we have distinguished between
priors observer can get from someone else (`the preparer', see discussion
of Eq. (11)) and from his own records. The `gut feeling' of this
author is that {\it all the mixtures are due to entanglement or correlations --
that they are all ultimately `improper'}. This is certainly possible if the Universe, as
a whole, is quantum. This is also suggested by quantum formalism, which
`refuses to recognize' (e.g., in the form of density matrices) any difference
between proper or improper mixtures.

Distinction between classical and quantum `missing information' is a related
issue. {\it Quantum discord} [41] seems to be a good way to measure some
aspects of the quantumness of information. And as there are states -- pointer
states -- that are effectively classical, the question arises as to whether
being ignorant of the relatively objective [15,3] state of the pointer
observable (that, as time goes on, is making more and more imprints on the
environment, `advertising' its states, so that they can be found out by many
without being perturbed) and being ignorant of a state of an isolated quantum
system differ in some way.

{\it Quantum Darwinism} [3,4,60] sheds a new light on these issue. According to
quantum Darwinism, classicality is an emergent property of certain observables
of a quantum Universe. It arises through selective proliferation of information
about them. Redundancy [3,4,6,60-63] is the measure of this classicality
-- observables are effectively classical when they have left many independently
accessible records in the rest of the Universe.  Approximate classicality 
arises when there are very many such records which can be independently
consulted. Such proliferation of information is enough to explain 
`objective classical reality' we experience. The idea of using redundancy 
to disitinguish between classical and quantum has some `prehistory' [6,61], and 
its role in the emergence of objectivity was brought up before [15]. Quantifying 
redundancy and exploring its consequences is an evolving subject
[3,4,60-63] which has recently led to new insights into the role of pointer
observables and into nature of ``objective existence" in the quantum universe.
Indeed, one can regard it as a modern embodiment of the ideas of Bohr [44] on
the role of amplification.

In light of quantum Darwinism the distinction between ignorance
about the future quantum and classical state (and, hence, quantum or classical
probabilities) can be understood as follows: When observer knows the 
preexisting state of a single, isolated quantum system, he may be ignorant of 
its future post-measurement state (and, hence, outcome of his about - to - be 
- carried - out measurement). This ignorance is quantum (or at least it is not classical) 
in the sense that the observer cannot discover what's already `out there' -- there is no 
`classical reality' that can be attributed to an isolated state of a quantum system. 

On the other hand, when there are many copies of the same information 
(about pointer states), then an initially ignorant observer will be able to 
deduce from the information in the environment which of the the stable 
pointer states of the system was responsible for the imprint. Moreover, he may
confirm his deductions obtained indirectly with a direct measurement (which can
be now designed as non-demolition -- observer has enough information to know
what to measure). In that situation the assumption that there was a preexisting
state of ${\cal S}$ that could be found out without being perturbed is (at least
in part) justified by the symptoms. 

This {\it relatively objective existence}
is all quantum theory has to offer to account for `classical reality', but this
seems to be enough. Quantum Darwinism's account of the emergence of classical
reality is in accord with the existential interpretation of quantum theory
[3,4,8,15]. Equally importantly, it is a good model for how we acquire most of
our information -- by intercepting a small fraction of the information present
in the photon environment. 


Envariance - based
definition of pointer states should be explored in much more detail, and its
consequences compared with the more traditional definitions introduced in the
studies of decoherence. In simple situations (e.g., idealized measurements,
Refs. [5-9]) there is no reason to expect any differences with the pointer
states selected by the predictability sieve. However, in realistic models
predictability sieve often leads to overcomplete sets of pointer states [3,15,39], 
and then it is not clear how should one go about
deducing pointer states from envariance.

Above list -- definition of systems, proper vs, improper mixtures, 
and ``pointer
states without decoherence'' are just the top three positions of a much longer
set of questions concerning {\it theoretical} implications of envariance.
However, over and above all of these items one should place a need for
a thorough {\it experimental} verification of envariance. To be sure,
envariance is a direct consequence of quantum theory, and quantum theory 
has been thoroughly tested. However, not all of
its consequences have been tested equally thorougly: Superposition principle
is perhaps the most frequently tested of the quantum principles. Tests of
entanglement are more difficult, but by now are also quite abundant. They have
focused on the violations of Bell's inequalities and, more recently, on various
applications of entanglement as a resource. These generally require 
global state
preparations but local measurements. Tests of envariance would similarily
require global preparations (of ${\cal SE}$ state), local manipulations
(e.g., swaps), but it would be very desirable to also have a global final
measurement to show that, following swap in the system, one may restore the
pre-swap global state of the whole with counterswap in the environment.

Untill recently, the combination of preparations, manipulations, and detections
required would have put experimental verification of envariance squarely in the
`gedanken' category, but recent progress in implementing quantum information
processing may place it well within the range of experimental possibilities.
Stakes are high: Envariance offers the chance to understand the ultimate origin
of probability in physics. The ease with which ignorance can be understood
in the quantum Universe and the difficulty of various classical approaches
combine to suggest that perhaps all probabilities in physics are fundamentally
quantum.

Envariance sheds a new, revealing light on the long suspected information -
theoretic role of quantum states. Moreover, it provides a new, physically
transparent, and deep foundation for the  understanding of the emergence of the
`classical reality' from the quantum substrate. In particular, envariance
provides an excellent example of the {\it epiontic} nature of quantum
states [3]: Quantum states share the role of describing what the observer
knows and what actually exists. In the classical realm these two functions are
cleanly separated. Their `quantum inseparability' was regarded as the source of
trouble for the interpretation of quantum theory. In the envariance - based
approach to probabilities it turns out to be a blessing in disguise --
it gives one an objective way to quantify ignorance, and it leads to Born's
rule for probabilities.

I would like to thank my colleagues Howard Barnum, Robin Blume-Kohout, 
Fernando Cucchietti, Harold Ollivier, and especially David Poulin (who provided
me with extensive comments on the manuscript) for stimulating discussions.

\section{Appendix A: Born's rule for continuous spectra}

The derivation of Born's rule we have put forward in the body of the paper
applies when the number of the participating Schmidt states is finite.
Here I shall extend first to the case of countably infinite number of states,
and then to the case of continuous spectra (e.g., derive $p(x)dx = |\psi(x)|^2dx$).
There are several ways to proceed. I shall present (briefly, and to some extent at
the expense of mathematical rigor) the proof that is physically most straightforward.

The case of infinitely many participating orthonormal Schmidt states (e.g., Eq. (2a),
but with $N=\infty$) can be systematically approximated with a sequence of finite 
but increasingly large $N_{\delta}$, chosen to be large enough to account for 
almost all likely alternatives. This can be done by splitting the Schmidt decomposition
into a sum of dominant contributions and a remainder:
$$ |\Psi_{\cal SEC}\rangle = \sum_{k=1}^{N_\delta} a_k |s_k\rangle |\varepsilon_k\rangle |c_k\rangle+
(\sum_{k=N_\delta+1}^{\infty} a_k |s_k\rangle |\varepsilon_k\rangle) |c_{N_\delta+1}\rangle$$
$$=\sum_{k=1}^{N_\delta} a_k |s_k\rangle |\varepsilon_k\rangle |c_k\rangle+
\delta |r_{N_{\delta}+1}\rangle |c_{N_\delta+1}\rangle \eqno(A1)$$
As $N_{\delta}$ increases, terms with the ``next largest'' $|a_k|$ are moved from the
unresolved remainder so that the absolute value of the coefficient $\delta$ in front of
the normalized $|r_{N_{\delta}+1}\rangle$ decreases. Using the previous
argument based on envariance one can readily see that:
$$ \forall_{k\leq N_{\delta}} \ \ p_k=p(s_k)=p(c_k) = |a_k|^2 \eqno(A2) $$
Moreover, the probability of the remainder is:
$$p(c_{N_\delta+1})=|\delta|^2 \eqno(A3)$$
It follows that Born's rule holds for every $s_k$.

The same conclusion can be reached in a slightly more roundabout way that 
involves conditional probabilities. Probability of the remainder is
$p(c_{N_\delta+1})=|\delta|^2$. Conditional probabilities of $s_k$ {\it given}
that $k\leq N_{\delta}$ are therefore:
$$ p_{k|k\leq N_{\delta}} = {{|a_k|^2} \over {1 - |\delta|^2}} $$
In the limit of vanishing $\delta$ this yields Born's rule for a sequence of two
measurements. The outcome of the first is -- in that limit -- certain: It establishes
that the unknown state is not a remainder. The second measurement is 
the ``high resolution'' followup.

The case of the continuous $\psi(x)$ can be treated by discretizing it. The most natural
strategy is to introduce a set of orthogonal basis functions that allow for a discrete
approximation of $\psi(x)$:
$$|\psi(x)\rangle \approx \sum_k \psi_k |k_\sqcap\rangle \ . \eqno(A4)$$
For instance, we can choose:
$$|k_{\sqcap}\rangle = 1 \ {\rm for} \ x \in [k \delta x, (k+1)\delta x), \ 0 \ {\rm otherwise}. \eqno(A5)$$
These functions are neither complete on the real axis, nor are they normalized:
$$ \langle k_{\sqcap} | k_{\sqcap}'\rangle = \delta x \times \delta_{kk'} \ . \eqno(A6)$$
Normalization can be achieved by dividing each $|k_{\sqcap}\rangle$ by $\sqrt {\delta x}$.
The coefficients $\psi_k$ are given by:
$$\psi_k = {1 \over  \delta x} \int_{k \delta x}^{(k+1) \delta x} dx~ \psi(x) =  \langle \psi(x) \rangle_k \ . \eqno(A7) $$
Now:
$$ |\psi (x) \rangle = |\Xi (x) \rangle + (|\psi(x)\rangle - |\Xi (x) \rangle) =  
\sum_k \psi_k |k_\sqcap\rangle + |r(x)\rangle \ . \eqno(A8)$$
Moreover,
$$ \langle \Xi (x)| r(x)\rangle = 0 \ , \eqno(A9)$$
as can be seen using Eq. (A7). This orthogonality of the discrete approximation 
to the state and the remainder is useful (but not essential)
to the discussion below. Furthermore:
$$ \langle \Xi (x)|\Xi (x) \rangle = \sum_k |\psi_k|^2 \delta x = 1 - |\delta|^2 \leq 1 \ . \eqno(A10)$$
For wave functions that are smooth on small scales in the limit $\delta x \rightarrow 0$ the norm
of the remainder vanishes, $\langle r(x)|r(x)\rangle = |\delta|^2 \rightarrow 0$.

One may wonder what happens when $\psi(x)$ is not sufficiently regular on small scales
to allow for the discrete approximation above to go through without complications. 
It is indeed possible to imagine, for example, fractal wavefunctions or situations 
where in addition to continuous $\psi(x)$ there are discrete points associated with 
a non-negligible contribution to the total probability. We shall bypass issues 
that arise in such cases. Their treatment is fairly straightforward, and has more 
to do with the theory of integration 
than with physics. Realistic wavefunctions tend to be sufficiently smooth on
small scales. For $\psi(x)$ finiteness of the total energy usually suffices to guarantee this.

We can now resume our derivation of Born's rule. Our aim is to calculate the
probability density associated with $\psi(x)$. In the limit of very small $\delta x$
$$ |\langle \psi(x) |\Xi (x) \rangle |^2=1-|\delta|^2 \rightarrow 1\ . \eqno(A11)$$
Therefore, $\psi(x)$ and $\Xi (x)$ have to yield the same probability density 
for all measurements. (In effect, we are using here again the assumption 
of continuity that was already invoked in section II).

So, the probabilities of detecting the system within various position intervals
can be inferred using envariance from the Schmidt decomposition:
$$|\Upsilon_{\cal SAE} \rangle = \sum_k |\psi_k| e^{i\phi_k} \sqrt{\delta x} 
{{|k_\sqcap\rangle} \over \sqrt{\delta x}}  |A_k\rangle |\varepsilon_k\rangle \ . \eqno(A12)$$
The set $\{{{|k_\sqcap\rangle} \over \sqrt{\delta x}}  \}$ is orthonormal (and, hence, 
can be regarded as `Schmidt'). Therefore, the complex Schmidt coefficients above
can be  used in the proof that is essentially the same as the one given in 
Sections II and V. We shall not restate it here in detail (although the reader may 
find it entertaining to re-think it in terms of approximate swapping of the sections 
of $\psi(x)$). The conclusion is inescapable: The probability of finding the apparatus 
in the state $|A_k\rangle$ (and, hence, probability of the system being found 
in the interval $x \in [k \delta x, (k+1)\delta x)$ is given by:
$$ p_k = |\psi_k|^2 \delta x \ . \eqno(A13)$$
It also follows that the probability of finding the system in 
the larger interval $[x_1,x_2)$ is:
$$p(x_1 \leq x < x_2)= \int_{x_1}^{x_2}dx |\psi(x)|^2 \ . \eqno(A14)$$
This establishes our premise. 

As we have already noted, we have ``cut corners'' and settled for a 
physically straightforward argument at the expense of the mathematical rigor.
We note that the basic structure of the argument can be refined, and that
mathematical rigor can be regained. For instance, there is no reason to 
use the same $\delta x$ everywhere, and one could improve convergence
properties of the discrete approximation by adapting the resolution of the
mesh that is better adapted to the form of $\psi(x)$. Moreover, the basis
states $|k_\sqcap \rangle$ are very artificial and violate the smoothness
assumption we have imposed on $\psi(x)$. They can be easily replace with 
more sophisticated orthonormal wavelets. 

Such mathematical improvements are beyond the scope of this work. They
help us however make an interesting physical point: Each such new discretization 
of $\psi(x)$ defines in effect a new measurement scheme, which will in turn 
impose its own definition of what exactly does it mean for the system to be found 
in a certain position interval. It is not essential to have a unique ``correct''
scheme. It is however important for all schemes that can be reasonably 
regarded as representing an approximate measurement of position should yield
compatible answers. In our case this is guaranteed, as in the limit of sufficient 
resolution all legitimate approximations of $\psi(x)$ are also clearly legitimate 
approximations of each other.   

\bigskip

\noindent{References}

\medskip

\noindent[1] M. Born, {\it Zeits. Phys.} {\bf 37}, 863 (1926).

\noindent[2] W. H. Zurek, {\it Phys. Rev. Lett.} {\bf 90}, 120404 (2003).

\noindent[3] W. H. Zurek, {\it Rev. Mod. Phys.} {\bf 75}, 715 (2003).

\noindent[4] W. H. Zurek, Quantum Darwinism and Envariance, 
quant-ph/0308163 (2003).

\noindent[5] W. H. Zurek, {\it Phys. Rev.} {\bf D24}, 1516 (1981).

\noindent[6] W. H. Zurek, {\it Phys. Rev.} {\bf D26}, 1862 (1982).

\noindent[7] E. Joos, H. D. Zeh, C. Kiefer, D. Giulini, J. Kupsch, 
I.-O. Stamatescu, {\it Decoherence and the Appearancs of 
a Classical World in Quantum Theory}, (Springer, 2003).

\noindent[8] W. H. Zurek, {\it Phil. Trans. Roy. Soc.} London Series A 
{\bf 356}, 1793 (1998).

\noindent[9] J.-P. Paz and W. H. Zurek, in {\it Coherent Atomic Matter Waves, 
Les Houches Lectures}, R. Kaiser, C. Westbrook, and F. David, eds.
(Springer, Berlin 2001).

\noindent[10] L. Landau, {\it Zeits. Phys.} {\bf 45}, 430 (1927). 

\noindent[11] M. A. Nielsen and I. L. Chuang, {\it Quantum Computation 
and Quantum Information} (Cambridge University Press, 2000).

\noindent[12] H. Everett, III, {\it Rev. Mod. Phys.} {\bf 29}, 454 (1957).

\noindent[13] H. Barnum, No-signalling-based version of Zurek's derivation 
of quantum probabilities: A note on ``Environment-assisted invariance,
entanglement, and probabilities in quantum physics, quant-ph/0312150
{\it Phys. Rev.} A, to appear (2004).

\noindent[14] W. H. Zurek, {\it Physics Today} {\bf 44}, 36 (1991); see also
an `update', quant-ph/0306072.

\noindent[15] W. H. Zurek, {\it Progr. Theor. Phys.} {\bf 89}, 281 (1993).

\noindent[16] M. Schlosshauer and A. Fine, On Zurek's derivation
 of the Born rule, quant-ph/0312058, {\it Found. Phys.}, submitted (2003).

\noindent[17] P. S. de Laplace {\it A Philosophical Essay on Probabilities},
English translation of the French original from 1820 
by F. W. Truscott and F. L. Emory (Dover, New York 1951).

\noindent[18] R. von Mises, {\it Probability, Statistics, and Truth} 
(McMillan, New York 1939).

\noindent[19] T. L. Fine, {\it Theories of Probability: 
An Examination of Foundations} (Academic Press, New York 1973).

\noindent[20] B. V. Gnedenko, {\it The Theory of Probability} 
(Chelsea, New York 1968).

\noindent[21] H. Everett III, {\it The Theory of the Universal Wave Function}, 
pp. 1-140 in B. S. DeWitt and N. Graham, in 
{\it The Many - Worlds Interpretation 
of Quantum Mechanics} (Princeton University Press, 1973).

\noindent[22] N. Graham, {\it The Measurement of Relative Frequency} 
pp. 229-253 
in B. S. DeWitt and N. Graham, in {\it The Many - Worlds Interpretation 
of Quantum Mechanics} (Princeton University Press, 1973).

\noindent[23] B. S. DeWitt, {\it Phys. Today} {\bf 23} 30 (1970); also 
in {\it Foundations of Quantum Mechanics}, B. d'Espagnat, ed.
(Academic Press, New York 1971), see also pp. 167-218
in B. S. DeWitt and N. Graham, in {\it The Many - Worlds Interpretation 
of Quantum Mechanics} (Princeton University Press, 1973).

\noindent[24] D. Deutsch, {\it Proc. Roy. Soc.} London Series A {\bf 455} 3129 (1999).

\noindent[25] R. Geroch, {\it No\^us} {\bf 18} 617 (1984).

\noindent[26] H. Stein, {\it No\^us} {\bf 18} 635 (1984).

\noindent[27] A. Kent, {\it Int. J. Mod. Phys.} {\bf A5}, 1745 (1990).

\noindent[28] E. J. Squires, {\it Phys. Lett.} {\bf A145}, 67 (1990).

\noindent[29] E. Joos, pp. 1-17 in {\it Decoherence: Theoretical, 
Experimental, nd  Conceptual Problems}, Ph. Blanchard, D. Giulini, 
E. Joos, C. Kiefer, and I.-O. Stamatescu, eds. (Springer, Berlin 2000).

\noindent[30] A. M. Gleason, {\it J. Math. Mech.} {\it bf 6}, 885 (1957).

\noindent[31] J. B. Hartle, {\it Am. J. Phys.} {\bf 36}, 704 (1968);

\noindent[32] E. Farhi, J. Goldstone, and S. Guttmann, {\it Ann. Phys.} 
(N. Y.) {\bf 24}, 118 (1989).

\noindent[33] D. Poulin, quant-ph/0403212; C. M. Caves and R. Shack, quant-ph/0409144.

\noindent[34] Y. Aharonov and B. Reznik, {\it Phys. Rev.} {\bf A65}, 
052116 (2003).


\noindent[35] D. Deutsch, {\it Int. J. Theor. Phys.} {\bf 24}, 1 (1985).

\noindent[36] D. Wallace, {\it Stud. Hist. Phil. Mod. Phys.} 
{\bf 34}, 415 (2003).

\noindent[37] J. A. Wheeler, pp 182-213 in {\it Quantum Theory and
Measurement} J. A. Wheeler and W. H. Zurek, eds., 
(Princeton University Press, 1983).

\noindent[38] H. D. Zeh, in {\it New Developments on Fundamental Problems 
in Quantum Mechanics}, M. Ferrero and A. van der Merwe, eds. 
(Kluwer, Dordrecht, 1997); quant-ph/9610014.

\noindent[39] W. H. Zurek, S. Habib, and J.-P. Paz, 
{\it Phys. Rev. Lett.} {\bf 70}, 1187 (1993).

\noindent[40] A. Elby and J. Bub, {\it Phys. Rev.} {\bf A49}, 4213 (1994).

\noindent[41] H. Ollivier and W. H. Zurek, {\it Phys. Rev. Lett.} 
{\bf 88}, 017901 (2002); 
W. H. Zurek, {\it Ann. der Physik} (Leipzig) {\bf 9}, 855 (2000);
{\it Phys. Rev.} {\bf A67}, 012320 (2003).

\noindent[42] R. Sikorski, {\it Boolean Algebras} (Springer, Berlin 1964).

\noindent[43] G. Birkhoff and J. von Neumann, {\it Ann Math.} 
{\bf 37}, 823 (1936).

\noindent[44] N. Bohr, {\it Nature} {\bf 121}, 580 (1928).

\noindent[45] R. I. G. Hughes, {\it The Structure and Interpretation 
of Quantum Mechanics} (Harvard University Press, Cambridge 1989).

\noindent[46] J. S. Bell, {\it Physics} {\bf 1}, 195 (1964).

\noindent[47] B. de Finetti, {\it Theory of Probability} 
(Wiley, New York, 1975).

\noindent[48] M. Li and P. Vit\'anyi, {\it An Introduction 
to Kolmogorov Complexity and
its Applications} (Springer, New York, 1993).

\noindent[49] C. Rovelli, {\it Int. J. Theor. Phys.} {\bf 35}, 1637 (1996).

\noindent[50] H. Jeffreys, {\it Theory of Probability} 
(Clarendon Press, Oxford 1961).

\noindent[51] H. Barnum, C. M. Caves, J. Finkelstein, C. A. Fuchs, and 
R. Schack, {\it Proc. Roy. Soc.} London {\bf A456}, 1175 (2000).
 
\noindent[52] S. Saunders, quant-ph/0211138.

\noindent[53] M. Schlosshauer, quant-ph/0312059.

\noindent[54] H. S. Leff and A. F. Rex, {\it Maxwell's Demon 2} 
(IOP Publishing, Bristol, 2003).

\noindent[55] L. Szilard, {\it Zeits. Phys.} {\bf 53}, 840 (1929);
R. Landauer, {\it IBM J. Res. Dev.} {\bf 5}, 183 (1961);
C. H. Bennett, {\it IBM J. Res. Dev.} {\bf 32} 16 (1988). 

\noindent[56] W. H. Zurek,  {\it Nature} {\bf 374}, 119 (1989);
{\it Phys. Rev.} {\bf A67}, 012320 (2003).

\noindent[57] A. Peres and W. H. Zurek, {\it Am J. Phys.} {\bf 50}, 807 (1982).

\noindent[58] B. d'Espagnat, {\it Phys. Lett} {\bf A282}, 133 (2000).

\noindent[59] G. Bacciagaluppi, in {\it The Stanford Encyclopedia 
of Philosophy}, E. N. Zalta, ed., 
on http://plato.stanford.edu/entries/qm-decoherence (2003).

\noindent[60] H. Ollivier, D. Poulin, and W. H. Zurek, quant-ph/0307229;
quant-ph/0408125.

\noindent[61] W. H. Zurek, p. 87 in {\it Quantum Optics, Experimental 
Gravitation, and Measurement Theory}, P. Meystre and M. O. Scully, eds.
(Plenum, New York, 1983).

\noindent[62] W. H. Zurek, {\it Annalen der Physik} 
(Leipzig) {\bf 9}, 855 (2000).

\noindent[63] R. Blume-Kohout and W. H. Zurek, quant-ph0408147; also, 
in preparation. 

\end{document}